\documentclass[traditabstract]{aa}

\usepackage{color}
\usepackage{tabularx}
\usepackage{booktabs}
\usepackage{amsmath}
\usepackage{txfonts}
\usepackage{graphicx}
\usepackage{xfrac}
\usepackage{xspace}
\usepackage{lscape}
\usepackage{multirow}
\usepackage{url}
\usepackage{caption}

\usepackage{natbib}
\usepackage[latin1]{inputenc}

\graphicspath{{./figures/}{./figures_old/}}

\bibpunct{(}{)}{;}{a}{}{,} 

\newcommand\cyg{\mbox{Cyg\,X-1}\xspace}

\newcommand\chandra{Chandra\xspace}

\newcommand\phiorb{\ensuremath{\phi_\mathrm{orb}}\xspace}
\newcommand{\hde}{HDE~226868\xspace}
\newcommand{\msun}{\ensuremath{M_{\odot}}\xspace}

\begin{document}

\title{Chandra X-ray spectroscopy of the focused wind in the
Cygnus~X-1 system} 
\subtitle{III. Dipping in the low/hard state} 
\author{Maria Hirsch\inst{1} \and
        Natalie~Hell\inst{2} \and 
        Victoria~Grinberg\inst{3} \and
        Ralf~Ballhausen\inst{1} \and
        Michael~A.~Nowak\inst{4} \and
        Katja~Pottschmidt\inst{5,6} \and
        Norbert~S.~Schulz\inst{7} \and
        Thomas~Dauser\inst{1} \and
        Manfred~Hanke\inst{1} \and
        Timothy~R.~Kallman\inst{6}\and
        Gregory~V.~Brown\inst{2} \and
        J\"orn~Wilms\inst{1}
}

\institute{Dr.~Karl Remeis-Sternwarte and Erlangen Centre for
  Astroparticle Physics, Universit\"at Erlangen-N\"urnberg, 
  Sternwartstr.~7, 96049 Bamberg, Germany
  \thanks{Maria.Hirsch@sternwarte.uni-erlangen.de} 
  \and Lawrence Livermore National Laboratory, 
       7000 East Ave., Livermore, CA 94550, USA 
  \and Institut f\"ur Astronomie und Astrophysik, Universit\"at T\"ubingen,
       Sand 1, 72076 T\"ubingen, Germany
  \and Department of Physics, Washington University in St.\ Louis,
       Campus Box 1105, One Brookings Drive, St.\ Louis, MO
       63130-4899, USA 
  \and CRESST, Department of Physics, 
       and Center for Space Science and Technology, 
       UMBC, Baltimore, MD 21250, US
  \and NASA Goddard Spaceflight Center, 8800 Greenbelt Rd, 
       Greenbelt, MD 20771, USA
  \and MIT Kavli Institute for Astrophysics and Space Research, 
       NE80-6077, 77 Mass. Ave., Cambridge, MA 02139, USA 
}

\date{Received 17 January 2019 / Accepted 22 April 2019}

\abstract{We present an analysis of three \chandra High Energy
  Transmission Gratings observations of the black hole binary
  \cyg/\hde at different orbital phases. The stellar wind that is
  powering the accretion in this system is characterized by
  temperature and density inhomogeneities including structures, or
  ``clumps'', of colder, more dense material embedded in the
  photoionized gas. As these clumps pass our line of sight, absorption
  dips appear in the light curve. We characterize the properties of
  the clumps through spectral changes during various dip stages.
  Comparing the silicon and sulfur absorption line regions
  (1.6--2.7\,keV $\equiv$ 7.7--4.6\,{\AA}) in four levels of varying
  column depth reveals the presence of lower ionization stages, i.e.,
  colder or denser material, in the deeper dip phases. The Doppler
  velocities of the lines are roughly consistent within each
  observation, varying with the respective orbital phase. This is
  consistent with the picture of a structure that consists of
  differently ionized material, in which shells of material facing the
  black hole shield the inner and back shells from the ionizing
  radiation. The variation of the Doppler velocities compared to a toy
  model of the stellar wind, however, does not allow us to pin down an
  exact location of the clump region in the system. This result, as
  well as the asymmetric shape of the observed lines, point at a
  picture of a complex wind structure.}

\keywords{
   accretion, accretion disks
-- stars: individual (HDE\,226868, \mbox{Cyg\,X-1})
-- stars: winds, outflows
-- techniques: spectroscopic
-- X-rays: binaries
}

\maketitle

\section{Introduction}\label{sect:intro}

Discovered during a balloon flight in 1964 \citep{bowyer65}, Cygnus
X-1 is one of the best studied black hole X-ray binaries. Based on
radio parallax data, the distance of the system was measured to be at
$1.86^{+0.12}_{-0.11}$\,kpc \citep{xiang11a,reid11}\footnote{New Gaia
  measurements appear to favor a slightly larger distance of
  $2.38^{+0.20}_{-0.17}$\,kpc, possibly due to systematic error caused
  by the high optical brightness of the system and/or an orbital
  wobble \citep{gandhi2018}.}. \cyg consists of a
$(14.8\pm 1.0)\,M_\odot$ black hole that accretes from the strong
stellar wind of the $(19.2 \pm 1.9)\,\msun$ supergiant O9.7\,Iab star
\hde
\citep{murdin1971,webster72,walborn73,herrero95,caballero09,orosz11}.
The star and the black hole are in a quasi-circular
\citep[eccentricity $e=0.018\pm 0.002$;][]{orosz11} 5.599829(16)\,d
orbit \citep{webster72,brocksopp99,gies03} with an inclination of
$i=27.1^\circ \pm 0.8^\circ$ \citep{orosz11}. This corresponds to a
separation between the center of mass of the star and the black hole
of only $42.2\,R_\odot$, or 2.5\,stellar radii. Combined with the high
mass-loss rate of \hde
\citep[${\sim}
10^{-6}\,\msun\,\mathrm{year}^{-1}$;][]{puls06,herrero95},
this small separation means that the black hole is continuously
accreting material from the stellar wind, making Cyg~X-1 one of the
few persistent black holes in our Galaxy.

With its high equivalent hydrogen column density, $N_{\mathrm{H}}$, of
around $6 \times 10^{21}\,\mathrm{cm}^{-2}$ \citep{dotani97, schulz02,
  miller02, hanke08}, \cyg is associated with an X-ray dust scattering
halo \citep[see][for the first analysis of this halo]{bode85}.
\citet{xiang11a} found a dust containing cloud in the interstellar
medium (ISM) at ${\sim}0.885\,D$, where $D$ is the distance of the
system, to be the region responsible for the scattering halo.

Already soon after the identification of the optical counterpart of
Cyg~X-1, X-ray light curves were found to show a strong orbital
modulation of the X-ray absorption column, $N_\mathrm{H}$, due to
absorption of X-rays from the black hole in the stellar wind
\citep{LiClark1974, RemillardCanizares1984, balucinska00,
  poutanen2008, miskovicova12a,grinberg2015}. Together with
observations of the orbital modulation of optical lines from \hde
\citep{giesbolton86a,giesbolton86b,gies03}, these phenomena led to the
picture of the stellar wind of \hde as a line driven wind or CAK wind
after \citet[][see also \citealt{friend82} and
\citealt{morton1967}]{castor75} with an asymptotic velocity of
$v_{\infty} \sim 2000\,\mathrm{km}\,\mathrm{s}^{-1}$
\citep{muijres12}. This wind is disturbed by the gravitational
potential of the black hole, which leads to a focusing of the wind
toward the black hole; see \citet{friend82} and \citet{giesbolton86b}
for early models and \citet{elmellah:2019} for a modern treatment of
this process. In addition, the wind is also affected by the X-rays
from the compact object: Strong orbital modulation of $N_\mathrm{H}$
is seen during the canonical hard state of the black hole, where the
X-ray spectrum is dominated by a Comptonized power law \citep[][and
references therein]{parker:15a,nowak11,wilms06}. The typical
bolometric luminosity of Cyg~X-1 in this state is around
$2\times 10^{37}\,\mathrm{erg}\,\mathrm{s}^{-1}$
\citep[e.g.,][]{wilms06,nowak:99a}, albeit with a large uncertainty
due to our lack of knowledge of the UV spectral shape. Only very
little $N_\mathrm{H}$ modulation is seen during the thermally
dominated X-ray soft state \citep{Wen1999,Boroson2010}, in which the
typical bolometric luminosity is at most a factor two higher than in
the hard state \citep[e.g.,][]{tomsick:2014,zhang:97b}. Optical
spectra show the stellar wind to be strongly photoionized during the
latter state \citep{gies03,gies08}.

\begin{figure*}
  \sidecaption
  \includegraphics[width=12cm]{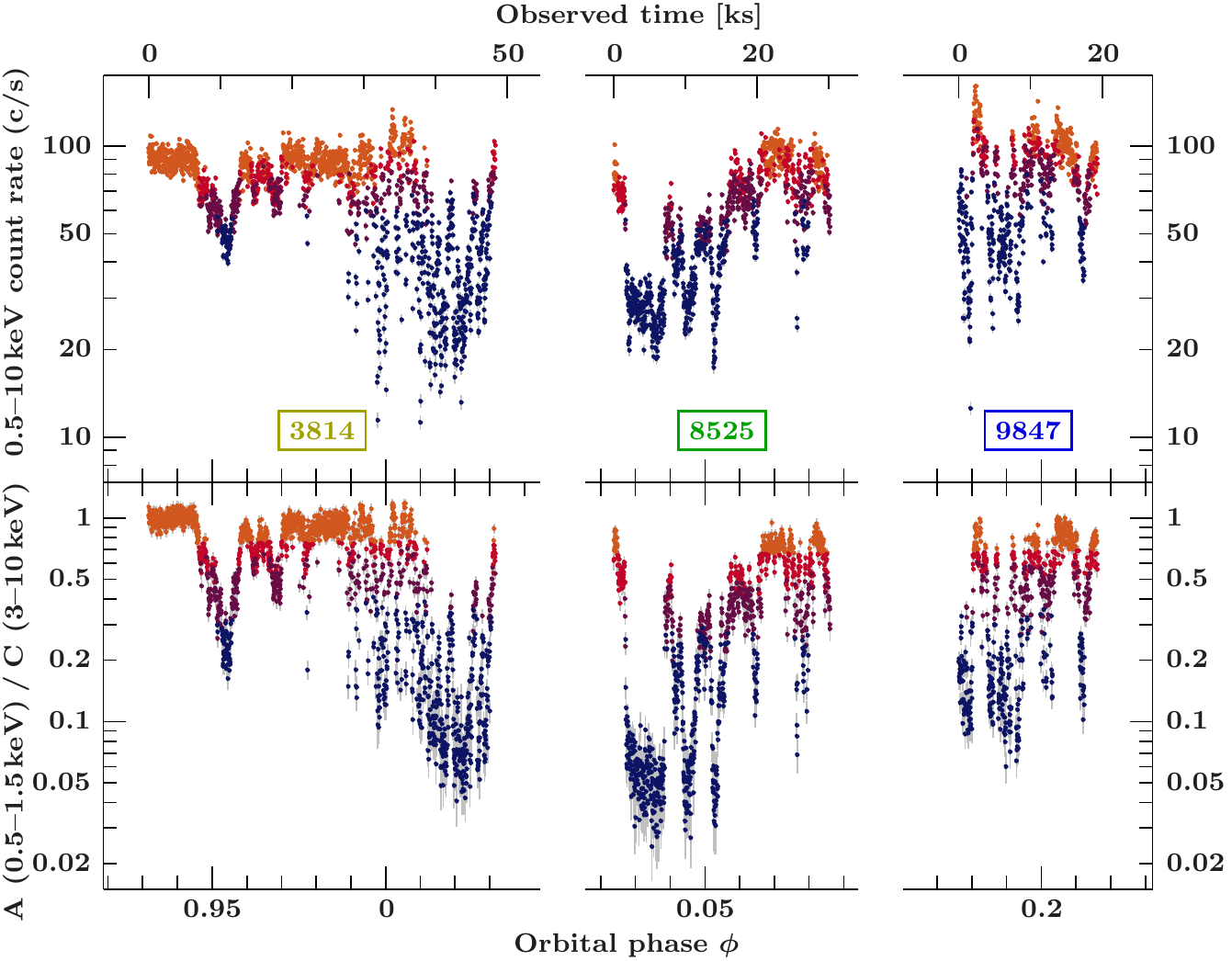}
  \caption{\emph{Top panel}: Light curves with a 25.5\,s time
    resolution of all three observations as a function of orbital
    phase. \emph{Bottom panel}: Corresponding hardness ratios. All
    observations were in the low/hard state Dips are strongest at
    $\phiorb\sim0.0$, become weaker at $\phiorb\sim0.2$ and
    $\phiorb\sim0.75$, and completely fade at $\phiorb\sim0.5$ (see
    also paper~II). Colors indicate the selected dipping stages (see
    right panel of Fig.~\ref{fig:color-color} for the color denotation
    and Sect.~\ref{sec:dataselection} for details on the data
    selection).}\label{fig:lightcurves}
\end{figure*}

Line driven winds are not expected to be smooth flows, but show strong
density perturbations or ``clumps'' \citep{owocki1988, feldmeier1997,
  puls06, puls2008,oskinova2012, sundqvist2013}. In X-ray binaries,
the density contrast could even be further enhanced by the interaction
between the wind and the strong X-rays from the compact object
\citep[][and references
therein]{blondin94,blondin95a,manousakis:11a,manousakis2015}. For
Vela~X-1 and Cyg~X-1 it has been estimated that more than 90\% of the
wind mass is contained in less than 10\% of the wind volume
\citep{sako99,rahoui11}. When the line of sight to the compact object
passes through one of these clumps, X-rays are absorbed by the
moderately ionized material in the clump, leading to a so-called
dipping event. This is also observed for other sources
\citep{hemphill2014,grinberg2017a}.  During the hard state of Cyg~X-1,
such short-term dipping events are observed predominantly during the
upper conjunction of the black hole, i.e., when the line of sight
passes through the densest region of the stellar wind and is most
likely to pass through a clump \citep{LiClark1974, mason1974,
  parsignault1976, pravdo1980, RemillardCanizares1984, kitamoto84,
  balucinska00, FengCui2002, poutanen2008,
  hanke09,miskovicova12a,grinberg2015}. The precise structure of the
clumps, i.e., their density and ionization structure, is unknown. Most
recent 2D simulations of such a stellar wind show a very complex
evolution of velocity and density structures with the formation of
characteristic small-scale clumps of various shapes embedded in areas
with lower density \citep{sundqvist2018}. Sundqvist et al. have found
a typical clump mass of $10^{17}$\,g and an average clump size of 1\%
of the stellar radius at a distance of two stellar radii.  These
results qualitatively confirm earlier theoretical models
\citep[e.g.,][]{oskinova2012,sundqvist2013} and observations
\citep[e.g.,][]{grinberg2015}. See also the review paper by
\citet{martinez2017}.

In this paper we present the first detailed high-resolution study of
absorption dips in \cyg using time-resolved X-ray spectroscopy during
dips observed with the High Energy Transmission Grating Spectrometer
(HETGS) on board Chandra. The analysis of a series of \chandra-HETGS
observations of the source taken during the low/hard state allows us
to investigate the dipping mechanism and to probe the ionization state
of the absorber directly. This is the third paper in a series devoted
to a study of \cyg using \chandra high-resolution spectroscopy. Our
previous analyses addressed the stellar wind in \cyg in the hard state
as seen outside the absorption dips, i.e., we analyzed the hot tenuous
phase of the wind. In the first paper of the series we investigated
the wind at $\phiorb\sim0$ \citep[][hereafter paper~I]{hanke09}. We
extended this work to the orbital modulation of the wind using non-dip
observations at $\phiorb\sim0.0$, $\sim$0.2, $\sim$0.5, and $\sim$0.75
\citep[][hereafter paper~II]{miskovicova12a}. These papers provide us
with a reference for the ``normal'' (non-dip) hard-state spectrum of
\cyg.

In this paper, we address the spectral changes during dipping
episodes. We discuss our data reduction and identification of the
dipping events in Sect.~\ref{sec:dips}. In
Sect.~\ref{sec:evolution} we study the spectral evolution of the silicon and
sulfur line regions from the non-dip to the deepest dip and we show that
absorption lines of lower ionized species appear as the line of sight
crosses through denser regions of the absorbing clump. During the
deepest dips the line of sight is fully blocked, revealing line
emission from the photoionized plasma around the black hole. 
We summarize our results in Sect.~\ref{sec:summary}.

\section{Data reduction}\label{sec:dips}

Out of the \chandra-HETGS observations discussed in paper~II for their
non-dip properties, we selected a subsample of three observations
(Table~\ref{tab:data}), namely ObsIDs~3814, 8525, and 9847, which show
distinct dipping episodes in their light curves
(Fig.~\ref{fig:lightcurves}, Sect.~\ref{sec:lightcurves}). As
discussed in paper~II, the other ObsIDs available were taken in the
intermediate and soft states and thus the material in the system was
in a different radiative environment. All observations selected for
further analysis of the dips were performed in timed exposure (TE)
mode. In this mode the usual frame time is 3.2\,s exposure before the
data are transferred into a frame store for readout. Cyg~X-1 is a bright
source even in the low/hard state. Therefore, for the discussed
observations only a half array of the chips (512 CCD rows) was read
out to minimize pileup by reducing the frame time to 1.7\,s. We
re-extracted the data using the Chandra Interactive Analysis of
Observations (CIAO) software version 4.6 with all parameters set to
default except for the HETG cross dispersion extraction width
(\texttt{width\_factor\_hetg} parameter in CIAO), which we reduced
from 35 to 10 for better coverage at the short
wavelength end.

For the light curves and spectral analysis we used the first order
spectra of the high and medium energy gratings \citep[HEG,
MEG;][]{canizares05}. Because of the excellent background
discrimination of the data extraction process (order sorting), any
remaining background is negligible compared to the bright source. We
therefore did not subtract any background from the final spectra. All
further data analysis was performed with the Interactive Spectral
Interpretation System version 1.6.2
\citep[ISIS;][]{Houck00,houck02,noble08a}.

\subsection{Light curves}\label{sec:lightcurves}

\begin{table}
\caption{\chandra-HETGS observations used in this paper}
\label{tab:data}
\centering
\begin{tabular}{llllll}
\hline
\hline
\multirow{3}{*}{ObsID} & \multicolumn{2}{c}{Start date} &
\multirow{3}{*}{Mode}& &
\multirow{3}{*}{$\phi_{\mathrm{orb}}$}\\
 & Date & MJD & & $T_{\mathrm{exp}}$ & \\
 & yyyy-mm-dd & & & [ks] & \\
\hline
3814 & 2003-04-19 & 52748 & TE/g & 48.3 & 0.93--0.03 \\
8525 & 2008-04-18 & 54574 & TE/g & 30.1 & 0.02--0.08 \\
9847 & 2008-04-19 & 54575 & TE/g & 19.3 & 0.17--0.21 \\
\hline
\end{tabular}\\
\tablefoot{TE/g: Timed exposure, graded;
  $T_{\mathrm{exp}}$: exposure time; $\Phi_{\mathrm{orb}}$: orbital
  phase according to the ephemeris of \citet{gies03}.}\\
\end{table}

\begin{figure*}
  \includegraphics[height=24\baselineskip]{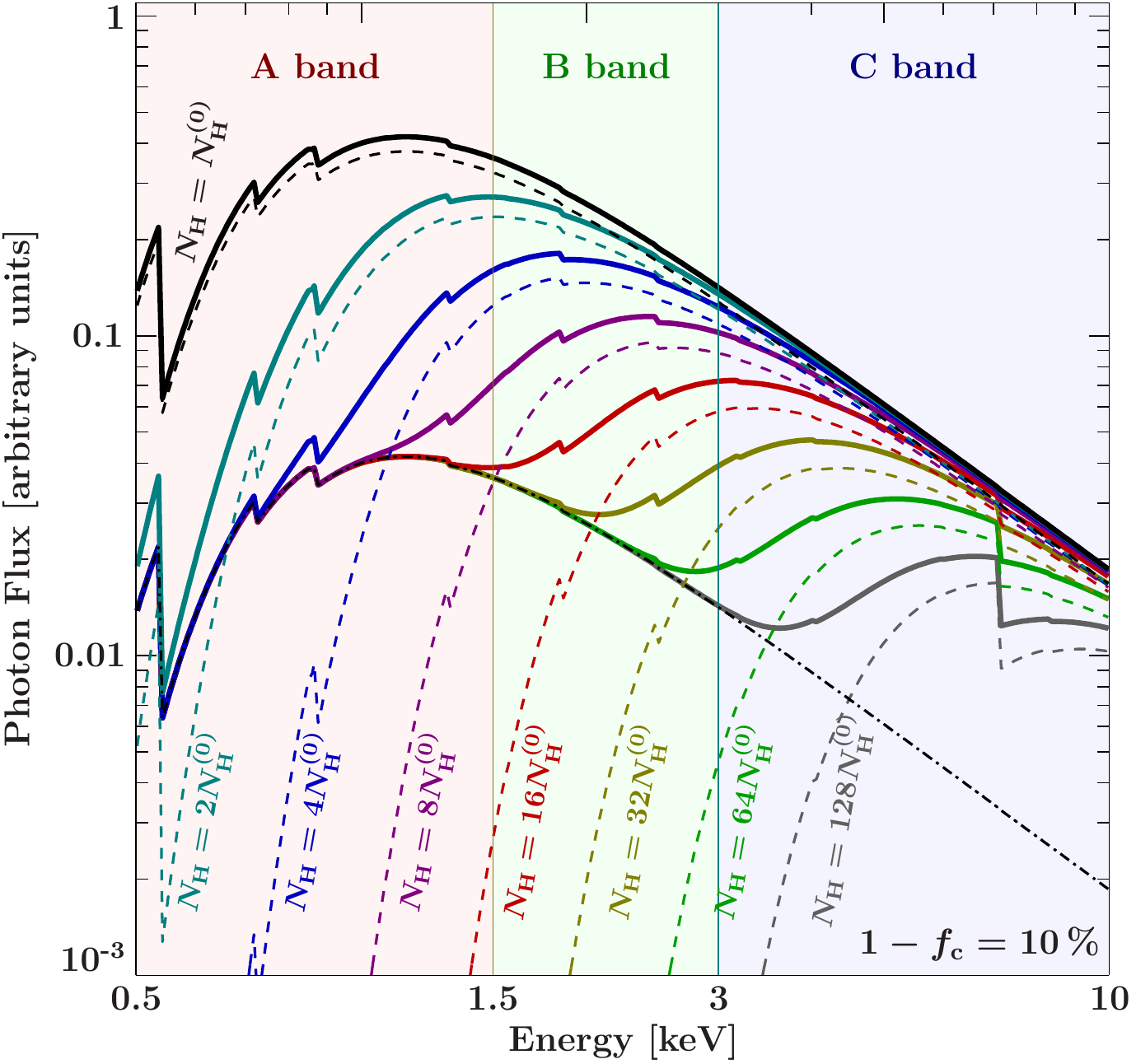}\hfill
  \includegraphics[height=24\baselineskip]{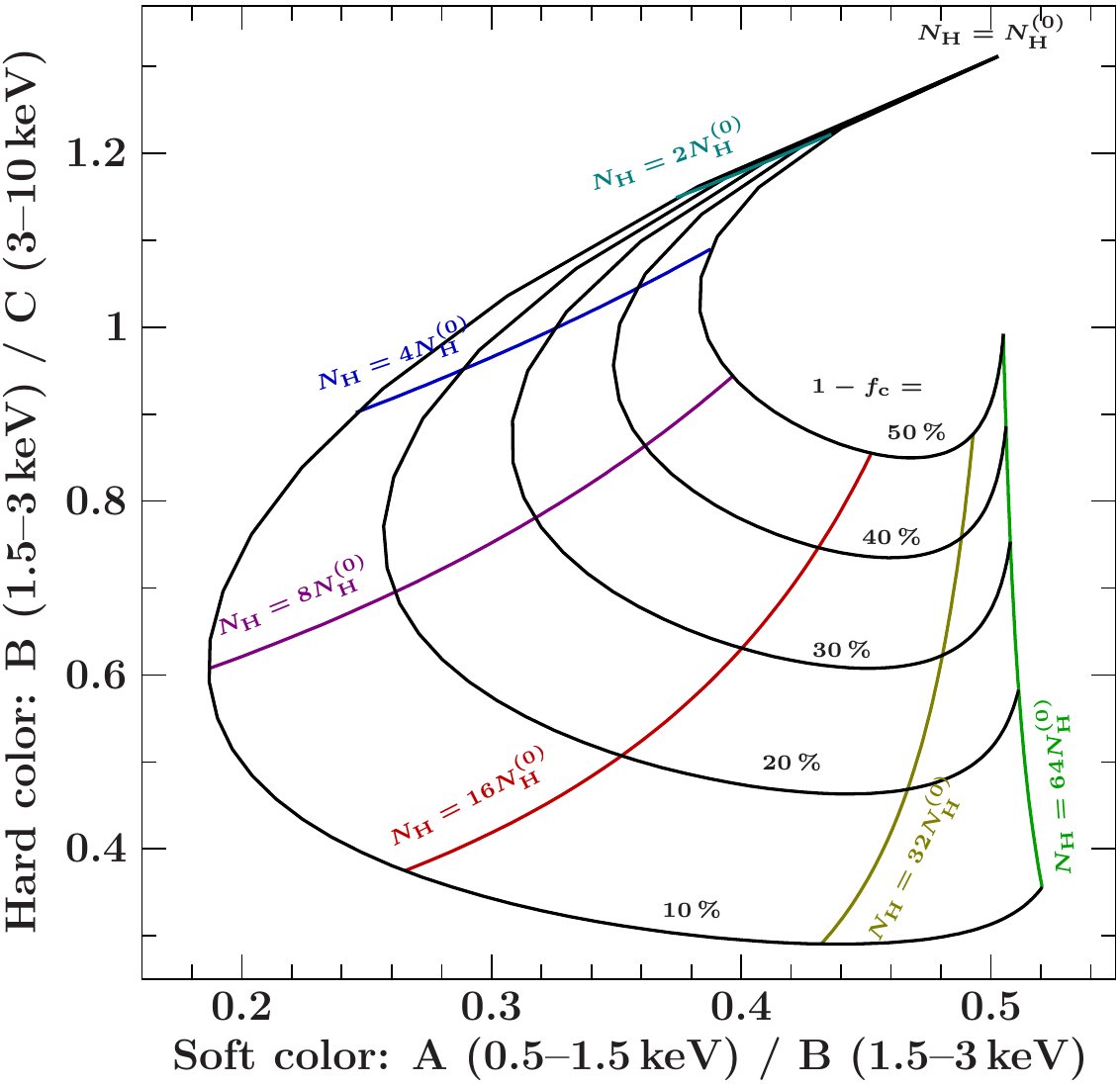}
  
  \caption{Left: Effect of partial absorption with increasing
    $N_\mathrm{H}$ at constant covering fraction with
    $N_\mathrm{H}^{(0)} = 5.4\times 10^{21}\,\mathrm{cm}^{-2}$ after
    paper~I. Right: Color-color diagram showing the tracks for
    different covering fractions and column densities of the second
    absorber. See text for details.}
  \label{fig:partial_covering}
\end{figure*}

The light curves of ObsIDs~3814 ($\sim$48\,ks) and 8525 ($\sim$30\,ks)
slightly overlap in orbital phase ($\phiorb=0.93$--0.03 and
$\phiorb=0.02$--0.08) and show a very similar morphology of strong
absorption dips. ObsID~9847 ($\sim$19\,ks) at $\phiorb=0.17$--0.21 was
taken within the same binary orbit as ObsID~8525; it contains
absorption dips as well, although they are less pronounced than at
$\phiorb\sim0$. Dips can last from several seconds to more than 10
minutes \citep{kitamoto84}. We extracted light curves with a 25.5\,s
resolution to uncover the shorter dips as well. For a detailed
description of the data set and a detailed analysis of the non-dip
spectrum of individual observations, we refer to papers~I
and~II. Paper~II also discusses ObsID~11044 ($\sim$30\,ks), which at
$\phiorb=0.48$--0.54 for the first time provides high-resolution
spectroscopy of \cyg during lower conjunction, i.e., when our line of
sight passes outside of the focused wind. As this light curve appears
to be virtually free of dipping, we do not discuss this light curve in
detail in this work.  \cyg was in a comparable hard state during
ObsIDs 3814, 8525, and 9847 according to the Rossi X-ray Timing
Explorer (RXTE) All Sky Monitor classification by \citet{grinberg13a};
see also the more detailed discussion of the source state in paper~II.
We do not use ObsID~3815 as it shows much less dipping than the other
hard state observations, which does not allow for a distinction
between different dip stages \citep[paper~II, see
also][]{miskovicova11}.

\subsection{Color-color diagrams}\label{sec:ccdiagrams}

As we discuss in more detail below, the dips are due to
absorption events caused by material in the line of sight to the
primary source of X-rays. Dipping events such as those seen in Cyg
X-1 are transient events characterized by quickly varying spectral
shape. This strong variability complicates the spectral analysis.
Ideally, we would want to study how the spectral shape -- photon
index, absorption/emission lines, and continuum absorption -- varies
with time, but the limited signal to noise of our observations renders
this impossible. We therefore have to resort to some kind of averaging
technique in which we extract spectra from time intervals where we
believe that the spectral shape is at least representative for a given
part of a dip. We find these time intervals by looking at the time
resolved spectral behavior of the source as represented in so-called
 color-color diagrams.

Figure~\ref{fig:partial_covering} shows how absorption affects an
observed primary power law continuum with $\Gamma=1.73$ \citep[typical
for Cyg X-1 in the hard state, e.g.,][]{grinberg13a} that is fully
covered by material with a fixed column density $N_\mathrm{H}^{(0)}$
(e.g., the outer parts of the stellar wind or absorption in the
ISM). For this paper, we set $N_\mathrm{H}^{(0)}$ to
the fit result of paper~I,
$N_\mathrm{H}^{(0)}=5.4\times 10^{21}\,\mathrm{cm}^{-2}$. In most
astrophysical sources, the structures responsible for dipping do not
cover the whole primary X-ray source, rather a partial coverer is
present in these systems that covers a fraction $f_\mathrm{c}$ of the
source with a column $N_\mathrm{H}$ (i.e., $1-f_\mathrm{c}$ remains
uncovered). A possible physical picture for such a partial coverer
would be an optically thick cloud that is smaller in (angular) size
than the X-ray source, or a cloud that passes very quickly over the
X-ray source, covering the source only for part of the integration
time. The observer therefore sees the sum of the uncovered spectrum
(dash-dotted line in Fig.~\ref{fig:partial_covering}) and the covered
spectrum (dashed lines). The summed spectrum is shown as solid lines
in Fig.~\ref{fig:partial_covering}. The left panel of
Fig.~\ref{fig:partial_covering} shows the observed spectral shapes
for a constant covering factor $f_\mathrm{c}=90\%$ and for varying
optical depths of the covering medium (up to $128N_\mathrm{H}^{(0)}$).
As $N_\mathrm{H}$ is increased, the partial coverer removes most of
the flux at soft energies and only the direct component remains
visible.

Because absorption events are typically of rather short duration, the
detailed spectral shape is often not directly observable. It is,
however, possible to characterize the spectral shape using X-ray
colors or hardness ratios \citep[see, e.g.,][for similar
approaches]{hanke08,nowak11}. In this work, we define the X-ray hardness ratio
as the ratio of the count rates in two energy bands. In order to be
consistent with paper~I we define the ratio such that its value
increases as the spectrum softens, so technically this ratio is a
``softness ratio''. We calculate hardness ratios using the fluxes
measured in three energy bands -- denoted as $A$ (0.5--1.5\,keV
$\equiv$ 24.8--8.27\,{\AA}), $B$ (1.5--3.0\,keV $\equiv$
8.27--4.13\,{\AA}), and $C$ (3.0--10.0\,keV $\equiv$
4.13--1.24\,{\AA}) with the ranges consistent with paper~II and the
previous work of \citet{nowak11} -- to characterize the spectral shape
in this way.

The right-hand panel of Fig.~\ref{fig:partial_covering} shows the
locus of such colors for the spectral shape discussed above and
several different covering fractions. For a constant $f_\mathrm{c}$,
a characteristic track in the color-color diagram becomes apparent: At
low $N_\mathrm{H}\sim N_\mathrm{H}^{(0)}$, the covered component
dominates the spectrum and the source is found in the top right of the
diagram. Since photoabsorption first influences the soft bands,
increasing $N_\mathrm{H}$ decreases both, $A/B$ and $B/C$, and the
source moves toward the bottom left of the diagram. At
intermediate $N_\mathrm{H}$, the A band is dominated by the (constant)
uncovered fraction, but increasing $N_\mathrm{H}$ still decreases $B$
and $C$. For this reason, $A/B$ starts to increase again, while $B/C$
continues to decrease, such that the track the source in the
color-color-diagram starts to turn toward the right. When the
contribution of the covered component to both A and B bands is almost
negligible, $A/B$ remains constant, while $B/C$ increases with
increasing $N_\mathrm{H}$. This behavior leads to a horizontal track
in which the source moves to the right in the color-color diagram.
Finally, for the largest $N_\mathrm{H}$ the entire covered fraction
is removed from the observable spectrum such that we expect the
X-ray color to move asymptotically back to its unabsorbed value.

\begin{figure*}
\centering \includegraphics[width=17cm]{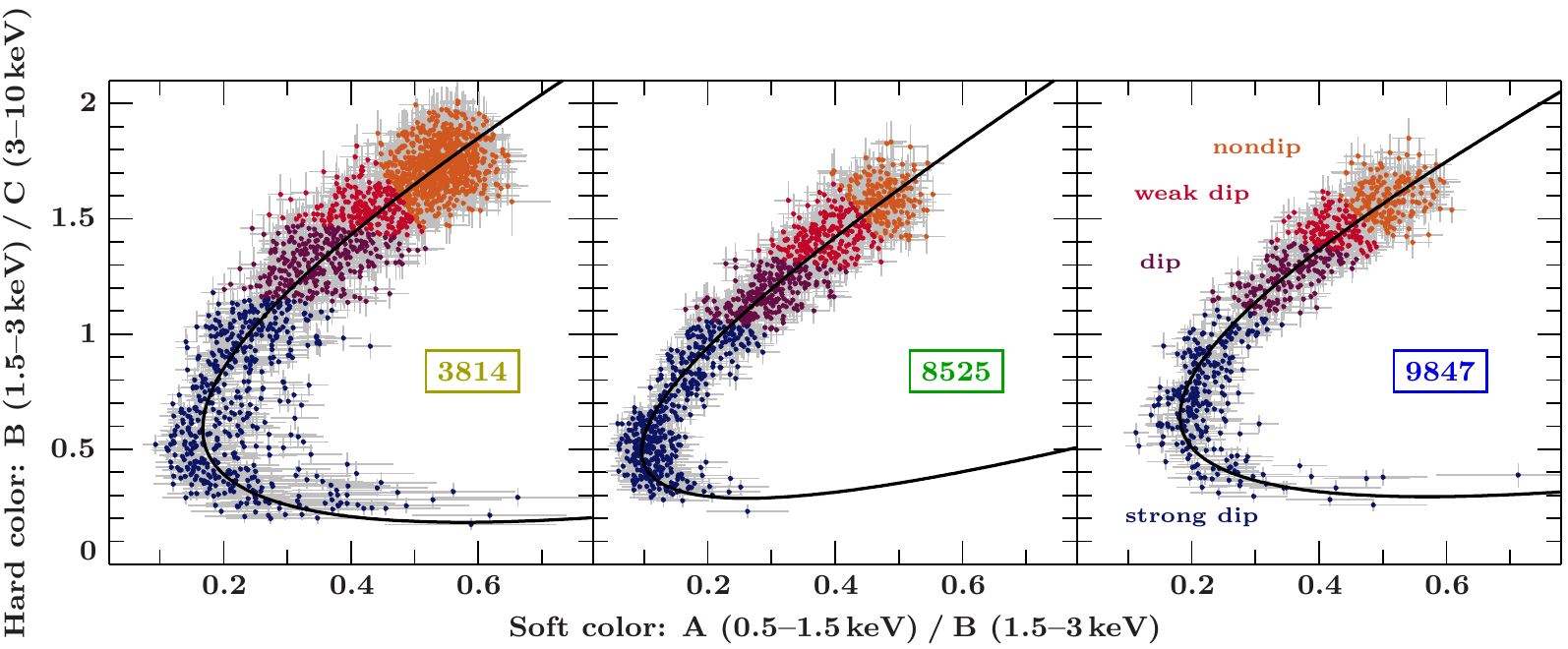}
\caption{Color-color diagrams for all three observations, showing
  hardness ratios calculated for soft, intermediate, and hard X-ray
  bands. As discussed in the text, the observed tracks are determined
  by absorption: following the track from the top right corner,
  representing the persistent flux, to the bottom left corner, showing
  dips, absorption increases and therefore both colors become harder.
  During the deepest dips, the softer color becomes softer again,
  while the hard color does not change. This behavior can be
  explained by partial covering (Sect.~\ref{sec:ccdiagrams}). The
  spectrum hardens considerably during these dips. The colors denote
  the four different dip stages (see right panel). The black line
  indicates our polynomial fit to the track; see
  Sect.~\ref{sec:dataselection} for details.}
\label{fig:color-color}
\end{figure*}

\subsection{Dip selection}\label{sec:dataselection}

We now apply the ideas discussed above to Cyg~X-1.
Figure~\ref{fig:color-color} shows the hardness ratios measured during
the three observations considered in this work. The data show a behavior
similar to that expected from partial covering: Data outside of dips
(orange data points) are barely absorbed; as the source enters deeper
dips (bluer colors), it follows a track that resembles the theoretical
tracks shown in Fig.~\ref{fig:partial_covering}, including a turning
point where the soft color becomes softer, while the hard color barely
changes.

To study the spectral shapes during different phases of the dips in
more detail, we extract data from time intervals that correspond to
various stages of dipping, i.e., phases of similar spectral shape. A
direct comparison of the theoretical tracks derived from the simple
partial covering model, however, shows deviations between the data and
the tracks. These deviations could be indicative of changes in both
the covering fraction and $N_\mathrm{H}$, ionization effects and
because the underlying X-ray continuum is more complex than a simple
absorbed power law. A detailed treatment of these effects, however,
would require us to fully understand the ionization structure of the
absorber, which is beyond the scope of this paper. We therefore
determine the shape of the track empirically by fitting an empirical
curve to the scatter plot. This curve is described through a
parameterized polynomial of second degree for each of the two colors
in the diagram. To find the polynomial coefficients, a $\chi^2$
minimization algorithm is used to optimize the shortest distance of
each data point to the curve.

To select data from the dips, we first remove the non-dip data as
defined in paper~II\footnote{The simpler non-dip selection in paper~II
  was done using only the ratio $A/C$.}. This ensures consistency with
the earlier results from paper~II. For the remaining data,
corresponding to various degrees of dipping, the goal is to find the
highest possible resolution in terms of number of dipping stages,
while maintaining a good enough signal-to-noise ratio to be able to
constrain the spectral fits well. To find this balance, we start out
with a large number of 12 segments, which are chosen such that each
slice contains roughly the same number of counts. Then, starting at
the deepest dipping stage (bottom right-hand corner of the color-color
diagram) we successively combine these small segments until the
signal-to-noise ratio of the resulting spectra is sufficient to detect
all possible lines of the Si and S series, but keep the number of
segments for this as low as possible to be able to distinguish a
greater number of dip stages. These combined segments constitute the
selection for the deepest dipping stage for further
analysis. Subsequently, the selection of the next dipping stages
follow the same approach until all 12 segments are sorted.

For all observations we obtain the best results by defining three
dipping stages in addition to the non-dip phase, each consisting
of four of the smaller segments. Consequently, the three stages each
have roughly the same number of counts within an observation. The four
dipping stages for each observation are classified as ``non-dip'',
``weak dip'', ``dip'', and ``strong dip'' (see Fig.~\ref{fig:color-color}
and Table\,\ref{tab:expo_counts} for the count rates and
exposure of each dip stage).

\begin{table}\renewcommand{\arraystretch}{1.2}

  \caption{Exposure time, $t_\mathrm{exp}$, and dip stage averaged
    count rates for the total observations and silicon (1.605 to
    2.045\,keV $\equiv$ 7.725 to 6.063\,{\AA}) and sulfur (2.295 to 
    2.7\,keV $\equiv$ 5.402 to 4.6\,{\AA}) bands for all observations.}
  \label{tab:expo_counts}
  \begin{tabular}{llrrrr}
     \hline
     \hline
      &  &                  &   \multicolumn{3}{c}{Counts} \\
      &  & $t_\mathrm{exp}$ & Total & Si & S \\
      &  &\multicolumn{1}{c}{[ks]} & \multicolumn{3}{c}{[$10^3\,\mathrm{counts}\,\mathrm{s}^{-1}$]} \\

    \hline
\multirow{4}{*}{\fbox{3814}  }
      & non-dip     & $21.096$ & 1838 & 37 & 92 \\
      & weak dip   &  $6.095$ &  445 & 10 & 26 \\
      & dip        &  $7.345$ &  461 & 13 & 31 \\
      & strong dip & $12.559$ &  458 & 21 & 49 \\
     \hline
\multirow{4}{*}{\fbox{8525}}
      & non-dip     &  $5.079$ &  450 & 11 & 22 \\
      & weak dip   &  $5.313$ &  394 & 11 & 26 \\
      & dip        &  $6.834$ &  406 & 13 & 28 \\
      & strong dip & $12.200$ &  401 & 20 & 41 \\
    \hline
\multirow{4}{*}{\fbox{9847}}
      & non-dip     &  $4.880$ &  499 & 12 & 24 \\
      & weak dip   &  $3.273$ &  294 &  8 & 16 \\
      & dip        &  $4.083$ &  307 &  9 & 19 \\
      & strong dip &  $6.618$ &  303 & 15 & 29 \\
    \hline
\end{tabular}
\end{table}

\section{Spectral evolution from non-dip to dip}\label{sec:evolution}
\begin{figure}
  \resizebox{\hsize}{!}{\includegraphics{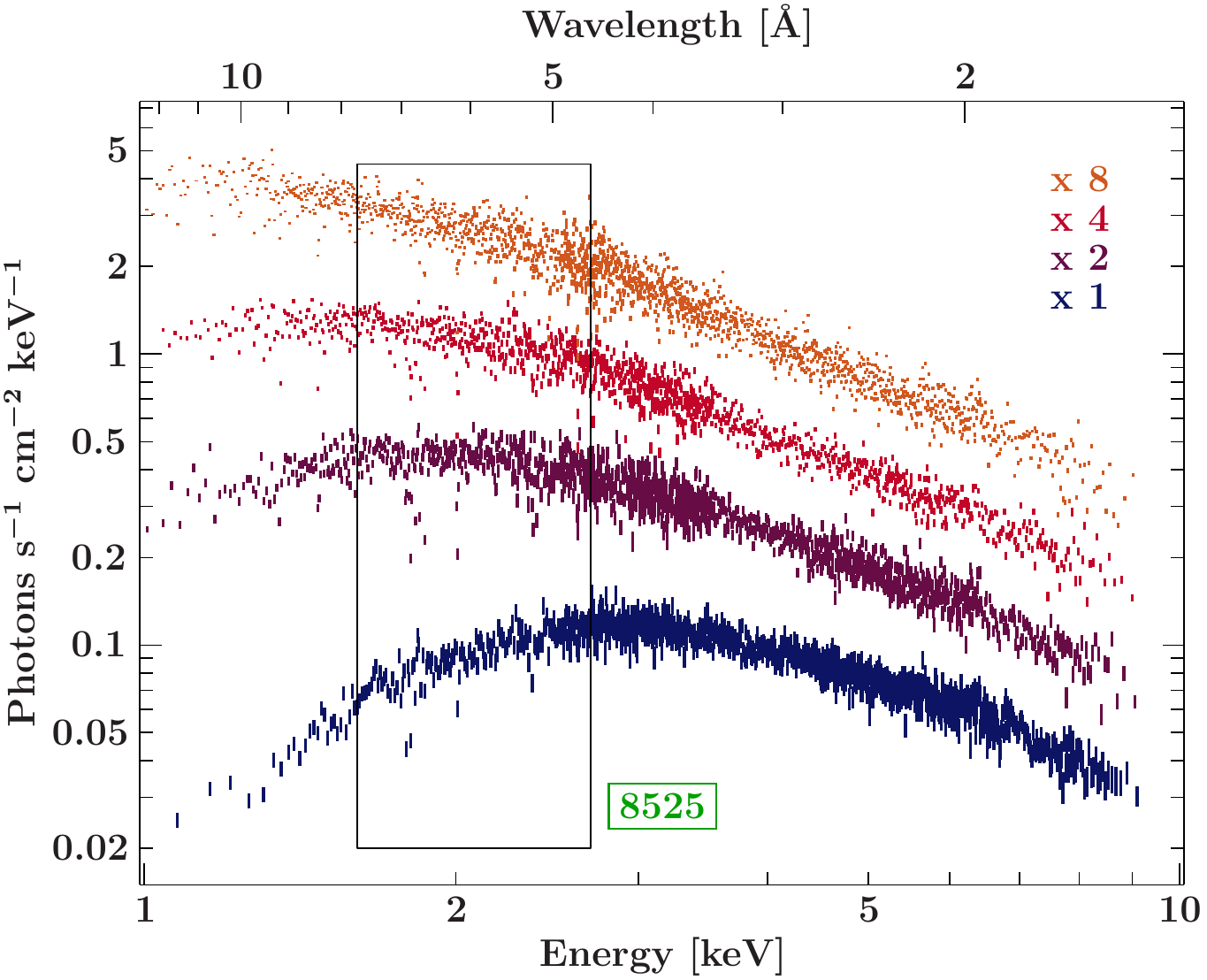}}
  \caption{Rebinned HEG spectra of the different dipping stages of
    ObsID 8525. For clarity, the non-dip spectrum (orange, $\mathrm{S/N}\ge13$)
    is scaled by a factor of 8, the weak dip spectrum (red, $\mathrm{S/N}\ge15$)
    by a factor of 4, and the dip spectrum (violet, $\mathrm{S/N}\ge15$) by a
    factor of 2. The strong dip spectrum (blue, $\mathrm{S/N}\ge17$) is not
    scaled. The box denotes the region containing the Si and S
    lines. The color scheme is the same as in
    Figs.~\ref{fig:lightcurves} and \ref{fig:color-color}.}
\label{fig:allspecs}
\end{figure}

Figure~\ref{fig:allspecs} shows the evolution of the spectrum during
dipping for ObsID~8525. In addition to the general change in spectral
shape due to photoelectric absorption, we see strong changes between
the individual dipping stages in the region between 1.6\,keV and
2.7\,keV (7.7--4.6\,{\AA}), where absorption lines of silicon and
sulfur ions are the most prominent spectral features (box in
Fig.~\ref{fig:allspecs}, see also \citealt{miskovicova11}). We
therefore concentrate on these Si and S lines.

\begin{figure*}
    \centering \includegraphics[width=17cm]{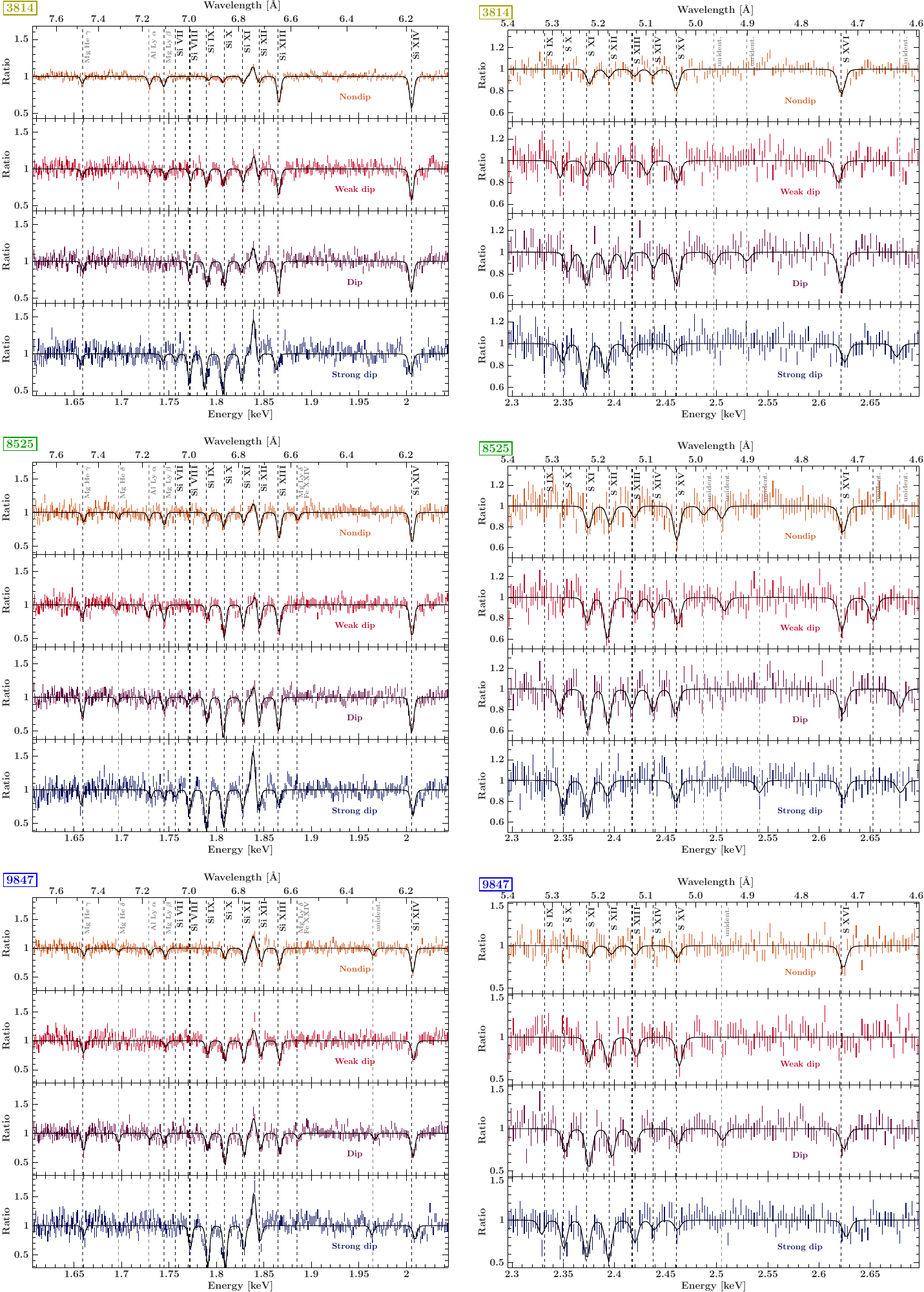}
    \caption{Evolution of silicon (left column) and sulfur (right
      column) lines with dipping stages for the three observations
      considered (from top to bottom). The panels show combined HEG
      and MEG $\pm 1^\mathrm{st}$ spectra relative to their
      respective continua. The solid black line shows the line fits. }
  \label{fig:spec_dipdepth}
\end{figure*}

\subsection{Si and S regions}\label{sec:lines}
For a detailed analysis of the Si and S absorption lines, we take into
account the $\pm$1 order HEG and MEG spectra. We combine both spectra
for the HEG and the MEG, and fit the spectra without rebinning them.
For clarity reasons, all figures show the combined HEG and MEG
spectra, where the higher resolution of the HEG is rebinned to the
lower resolution of the MEG. This combination of HEG and MEG, however,
only applies to the display of the data, not to the actual fitting.

As the emphasis of this paper is on the behavior of the Si and S
lines, we do not attempt to model the broadband Chandra continuum or
the continuum absorption, but rather describe the local spectra in the
Si (1.605--2.045\,keV $\equiv$ 7.725--6.063\,{\AA}) and the S
(2.295--2.7\,keV $\equiv$ 5.402--4.6\,{\AA}) regions. These regions
are narrow enough in energy that the curvature due to the absorption
is negligible. We can therefore describe the local absorbed continuum
by a simple power law; obviously, the photon indices of the continuum
are different in both bands.

A further complication is that as a consequence of the source brightness the
non-dip, weak dip, and to a lesser extent also the dip spectra are
affected by slight pileup. The major effect that pileup has on our
narrowband data is a change in count rate, which is seen as a change
in the relative flux normalization of both spectra, as pileup affects
the MEG more strongly than the HEG because of the significantly lower
spectral resolution\footnote{See `The Chandra ABC Guide to Pileup',
  CXC, 2010,
  \url{http://cxc.harvard.edu/ciao/download/doc/pileup_abc.pdf}}. We
can compensate for this effect by including a multiplicative, detector
dependent constant in the spectral modeling. Because of the narrow energy
bands considered here, a further pileup correction is not necessary.
The pileup fraction is a slowly changing function of energy such that
it does not appreciably affect the equivalent width of the lines
considered in this work; because of the nature of X-ray gratings, line energies
are not affected. See \citet{hanke09} for a discussion of pileup
effects in the non-dip spectrum of Cyg~X-1.

The local continuum being accounted for (see
Table\,\ref{tab:continuum} for parameters), we model the absorption
lines in both regions using additive Gaussian line profiles
(Fig.~\ref{fig:spec_dipdepth}). These profiles give a slightly better
description of the line shapes than Voigt profiles. For the absorption
lines, the line width $\sigma$ was frozen at 1.2\,eV (silicon) and
1.8\,eV (sulfur), well below the detector resolution. Lines were
identified using energies from \citet{hell16}, \citet{porquet10a}, and
from AtomDB, version 2.0.0\footnote{see \url{http://www.atomdb.org};
  line energies without a reference refer to AtomDB.}. We find blended
line complexes for L-shell ions, He\,w lines for He-like ions, and
Ly\,$\alpha$ lines for H-like ions of both silicon and sulfur. In
addition to the K\,$\alpha$ absorption line series, there is the
forbidden He-like \ion{Si}{xiii}\,z line in emission
\citep[1.8394\,keV $\equiv$ 6.7405\,{\AA},
$1\mathrm{s}2\mathrm{s}\,{}^3\mathrm{S}_1 \rightarrow
1\mathrm{s}^2\,{}^1\mathrm{S}_0$;][]{porquet10a};
this line is sometimes also called the He-like \ion{Si}{xiii} f line.
We note that there is a degeneracy between the Be-like \ion{Si}{xi}
and Li-like \ion{Si}{xii} absorption lines at 1.8275\,keV $\equiv$
6.7844\,{{\AA}} and 1.8450\,keV $\equiv$ 6.7200\,{\AA} \citep[both
energies from][]{hell16} and the forbidden He-like \ion{Si}{xiii}\,z
emission line. This increases the uncertainty of the respective fit
values. Finally, in the strong dip spectrum of ObsID 9847, there are
indications for a O-like \ion{S}{ix} K\,$\alpha$ line, which appears
once the HEG data are rebinned to match the (unbinned) MEG grid and
both are combined for display. This line is too weak for us to be able
to constrain its parameters such that we only claim a tentative
detection of O-like \ion{S}{ix}. The values for central line energies
and equivalent widths of all detected silicon and sulfur absorption
lines can be found in Tables~\ref{tab:energies} and~\ref{tab:eqws}. We
note that the lower ionization stages (N-like and O-like ions) only
appear deeper in the dips. This hints toward a shell-like ionization
structure of the clumps with a hot, highly ionized surface and cooler
and less ionized parts in the core.

In addition to these lines from S and Si we also find an
\ion{Al}{xiii} Ly\,$\alpha$\footnote{Lines denoted as part of the
  Lyman series typically have two unresolved components with the
  configuration of
  $1\mathrm{s}_{1/2}\,{}^2\mathrm{S}_{1/2} \text{--}
  \mathrm{np}_{3/2}\,{}^2\mathrm{P}_{3/2}$
  and
  $1\mathrm{s}_{1/2}\,{}^2\mathrm{S}_{1/2} \text{--}
  \mathrm{np}_{1/2}\,{}^2\mathrm{P}_{1/2}$,
  respectively. The labels $\alpha$, $\beta$, $\gamma$ etc. represent
  the principal quantum number $\mathrm{n}=2,3,4,\ldots$. The line
  energy given in the text corresponds to the mean energy of the components,
  weighted with the respective statistical weight (2:1).} absorption
line (1.7285\,keV $\equiv$ 7.1729\,{\AA}), Mg He\,$\gamma$
(1.6591\,keV $\equiv$ 7.4730\,{\AA},
$1\mathrm{s}^2\,{}^1\mathrm{S}_0 \text{--}
1\mathrm{s}4\mathrm{p}\,{}^1\mathrm{P}_1$)
and He\,$\delta$ (1.6961\,keV $\equiv$ 7.3100\,{\AA},
$1\mathrm{s}^2\,{}^1\mathrm{S}_0 \text{--}
1\mathrm{s}5\mathrm{p}\,{}^1\mathrm{P}_1$)
absorption lines, and we detect \ion{Mg}{xii}\,Ly\,$\beta$
(1.7447\,keV $\equiv$ 7.1063\,{\AA}) and, in some cases, a hint of Mg
Ly\,$\delta$ (1.8843\,keV $\equiv$ 6.5799\,{\AA}) blended with an
\ion{Fe}{xxiv} transition (1.8851\,keV $\equiv$ 5.5771\,{\AA},
$1\mathrm{s}^{2}2\mathrm{s}\,{}^2\mathrm{S}_{1/2} \text{--}
1\mathrm{s}^{2}7\mathrm{p}\,{}^2\mathrm{P}_{3/2}$).
These lines imply the presence of Mg\,Ly\,$\gamma$ at 1.840\,keV
$\equiv$ 6.738\,{\AA}, i.e., blended with He-like Si\,z. We expect
Mg\,Ly\,$\gamma$ to be very weak but it may slightly contaminate our
measurements of the Si\,He\,z and the Li-like \ion{Si}{xii} line, thus
further contributing to the uncertainty of the fit values.

\begin{figure*}
  \sidecaption
    \includegraphics[width=12cm]{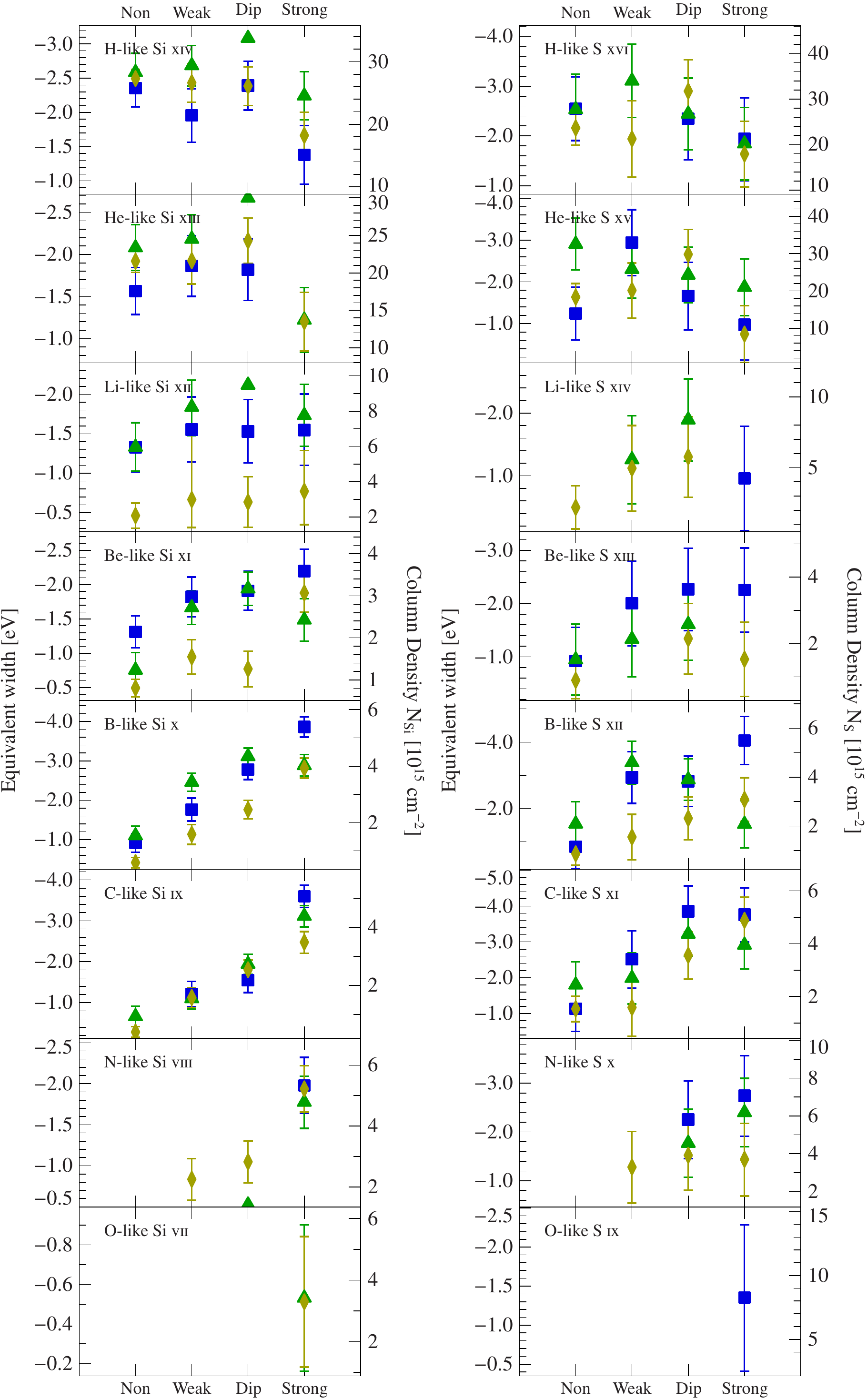}
    \caption{Column densities for each dip stage derived from the
      respective equivalent width for all Si and S ion absorption
      lines detected in ObsIDs 3814 (yellow diamonds), 8525 (green
      triangles), and 9847 (blue squares). Negative values of
      the equivalent width correspond to absorption lines. The column
      densities of lower charged ions increase with increasing dip
      depth, whereas column densities of highly charged ions decrease
      with dip level. Be-like \ion{Si}{xi} and Li-like
      \ion{Si}{xii} lines are contaminated by the Si\,He\,z emission
      line. } \label{fig:dipstage_column}
\end{figure*}

\subsubsection{Equivalent widths or line strengths and column densities}\label{sec:eqw}
The equivalent widths of the silicon and sulfur lines vary around
1\,eV and have rather large error bars for the weaker lines
(Table~\ref{tab:eqws}). We note that the He-like \ion{Si}{xiii}\,z
emission line additionally increases the uncertainties of the
\ion{Si}{xi} and \ion{Si}{xii} lines. Comparing the lines in the
different dipping stages reveals that the line strengths of the low
charge states -- and thus the respective column densities -- increase
with dipping, while those of the higher charge states decrease
(Fig.~\ref{fig:dipstage_column}). This shift in charge balance alone
is already strong evidence supporting the theory of a highly
structured medium where clumps of colder, denser material are
embedded in a highly ionized plasma \citep[][and references
therein]{oskinova2012, sundqvist2013}.

For optically thin absorption lines, the equivalent width directly
translates into the column density of the parent ion, $N_i$. In this
section and in the following we use the convention that the equivalent
width of absorption lines is negative and that of emission lines is
positive.  We derive the column densities for all Si and S ions
visible in our spectra from (paper~I, Eq.~11)
\begin{equation}
N_i = \frac{m_\mathrm{e}c^2 |W_\lambda|}{\pi e^2 f_{ij} \lambda_0^2} = \frac{9.11\times
10^{15}\,\mathrm{cm}^{-2}}{f_{ij}}\,\frac{|W_E|}{\mathrm{eV}},
\end{equation}
where $W_\lambda$ and $W_E$ are the equivalent width in wavelength
and in energy space, and where $f_{ij}$ is the oscillator strength,
taken as the sum of all transitions blending into the absorption
line of the respective ion. We use the compilation of
\citet{verner96table} for H-like ions, and the compilation of
\citet{palmeri08a} for all other ionization stages. These oscillator
strengths include Auger damping, which is important at the densities
expected. All other symbols have their usual meanings.

\begin{table*}\renewcommand{\arraystretch}{1.2}
  \caption{Column densities as calculated from the silicon line
    equivalent widths.}
  \label{tab:N_si}
\resizebox{\textwidth}{!}{%
  \begin{tabular}{lllllllllllll}
     \hline
     \hline
        & & $N_{\text{H-like}}$ & $N_{\text{He-like}}$ & $N_{\text{Li-like}}$ &
                                      $N_{\text{Be-like}}$ & $N_{\text{B-like}}$ &  $N_{\text{C-like}}$ & $N_{\text{N-like}}$ &
                                      $N_{\text{O-like}}$ & $N_{\mathrm{Si}}$ & $N_{\mathrm{H, Wilms}}$ & $N_{\mathrm{H, Herrero}}^{\dagger}$ \\
                                 &   &
                                 \multicolumn{8}{c}{[$10^{15}\,\mathrm{cm}^{-2}$]}
                                 & [$10^{15}\,\mathrm{cm}^{-2}$] &
                                 \small [$10^{21}\,\mathrm{cm}^{-2}$]
                                                                              &
                                                                                \small [$10^{21}\,\mathrm{cm}^{-2}$] \\
    \hline
    \multirow{4}{*}{\fbox{3814}} 
       & non-dip     & $54.6^{+3.0}_{-3.1}$ & $21.6\pm1.6$ & $2.1\pm0.8$         & $0.81\pm0.21$ & $0.59\pm0.19$ & $0.40\pm0.19$ & --          & --          & $80\pm7$  & $2.26\pm0.20$ & $0.89\pm0.08$ \\
       & weak dip   & $53\pm7$             & $22\pm4$     & $3.0^{+4.1}_{-1.6}$ & $1.5\pm0.5$   & $1.6\pm0.4$   & $1.6\pm0.4$   & $2.2\pm0.7$ & --          & $85\pm15$ & $2.4\pm0.5$   & $0.95\pm0.17$ \\
       & dip        & $52\pm7$             & $24\pm4$     & $2.8\pm1.5$         & $1.3\pm0.5$   & $2.5\pm0.4$   & $2.5\pm0.4$   & $2.8\pm0.7$ & --          & $88\pm15$ & $2.5\pm0.4$   & $0.99\pm0.16$ \\
       & strong dip & $36\pm8$             & $13\pm4$     & $3.5^{+2.4}_{-1.9}$ & $3.1\pm0.5$   & $3.9\pm0.4$   & $3.5\pm0.4$   & $5.2\pm0.8$ & $3.3\pm2.2$ & $72\pm18$ & $2.0\pm0.6$   & $0.81\pm0.21$ \\
     \hline
    \multirow{4}{*}{\fbox{8525}} 
       & non-dip     & $57\pm7$ & $23\pm4$ & $6.0\pm1.4$         &  $1.2\pm0.5$        & $1.5\pm0.4$            & $0.9\pm0.4$ & --          & --          & $90\pm15$ & $2.5\pm0.4$ & $1.00\pm0.16$ \\
       & weak dip   & $59\pm7$ & $24\pm4$ & $8.2^{+1.6}_{-1.7}$ & $2.7\pm0.5$         & $3.4\pm0.4$            & $1.5\pm0.4$ & --          & --          & $99\pm15$ & $2.8\pm0.5$ & $1.11\pm0.17$ \\
       & dip        & $68\pm7$ & $29\pm4$ & $9.5^{+1.5}_{-1.4}$ & $3.2^{+0.4}_{-0.5}$ & $4.35^{+0.30}_{-0.31}$ & $2.7\pm0.4$ & $1.4\pm0.8$ & --          & $119\pm15$ & $3.3\pm0.4$ & $1.32\pm0.16$ \\
       & strong dip & $49\pm8$ & $13\pm5$ & $7.8\pm1.8$         & $2.4\pm0.6$         & $4.0\pm0.4$            & $4.4\pm0.4$ & $4.8\pm0.9$ & $3.4\pm2.4$ & $90\pm19$ & $2.5\pm0.6$   & $1.00\pm0.21$ \\
    \hline
    \multirow{4}{*}{\fbox{9847}} 
       & non-dip     & $52\pm6$     & $18\pm4$ & $5.9\pm1.4$         & $2.1\pm0.4$ & $1.3\pm0.4$ & --          & --          & -- & $78\pm14$ & $2.2\pm0.4$ & $0.88\pm0.16$ \\
       & weak dip   & $43\pm9$     & $21\pm5$ & $7.0\pm1.9$         & $3.0\pm0.5$ & $2.5\pm0.5$ & $1.7\pm0.5$ & --          & -- & $78\pm20$ & $2.2\pm0.6$   & $0.87\pm0.22$ \\
       & dip        & $52\pm8$     & $20\pm5$ & $6.8^{+1.9}_{-1.8}$ & $3.1\pm0.5$ & $3.9\pm0.4$ & $2.2\pm0.5$ & --          & -- & $89\pm19$ & $2.5\pm0.6$   & $0.99\pm0.21$ \\
       & strong dip & $30\pm10$    & --       & $6.9\pm2.1$         & $3.6\pm0.6$ & $5.4\pm0.4$ & $5.1\pm0.4$ & $5.3\pm1.0$ & -- & $57\pm20$ & $1.6\pm0.6$ & $0.63\pm0.22$ \\
    \hline
\end{tabular}}\\%
{{}$^{\dagger}$\,\citet{herrero95} provide a scaling factor for the
  observed overabundance of helium, which we apply to the solar
  abundances of \citet{wilms00}. See text for further details.}
\end{table*}

\begin{table*}
  \caption{Column densities as calculated from the sulfur line
    equivalent widths.}
  \label{tab:N_s}
  \resizebox{\textwidth}{!}{%
  \begin{tabular}{lllllllllllll}
     \hline
     \hline
      &  & $N_{\text{H-like}}$ & $N_{\text{He-like}}$ & $N_{\text{Li-like}}$ &
                                      $N_{\text{Be-like}}$ & $N_{\text{B-like}}$ &  $N_{\text{C-like}}$ & $N_{\text{N-like}}$ &
                                      $N_{\text{O-like}}$ & $N_{\mathrm{S}}$ & $N_{\mathrm{H, Wilms}}$ & $N_{\mathrm{H, Herrero}}^\dagger$ \\
                                 &  &
                                 \multicolumn{8}{c}{[$10^{15}\,\mathrm{cm}^{-2}$]}
                                 & [$10^{15}\,\mathrm{cm}^{-2}$] &
                                 [$10^{21}\,\mathrm{cm}^{-2}$] & [$10^{21}\,\mathrm{cm}^{-2}$] \\
    \hline
    \multirow{4}{*}{\fbox{3814}} 
      & non-dip     & $47\pm8$ & $18\pm4$ & $2.2\pm1.6$ & $0.9\pm0.6$ & $0.9\pm0.5$         & $1.5\pm0.5$         & --                  & -- & $71\pm18$ & $2.0\pm0.5$ & $0.79\pm0.19$ \\
      & weak dip   & $43\pm17$ & $20\pm8$ & $5\pm4$     & --          & $1.6\pm1.0$         & $1.6\pm1.1$         & $3.3^{+1.9}_{-2.0}$ & -- & $70\pm40$ & $2.1\pm1.1$ & $0.8\pm0.5$ \\
      & dip        & $64\pm14$ & $30\pm7$ & $5.8\pm2.9$ & $2.2\pm1.1$ & $2.3\pm0.9$         & $3.6^{+0.9}_{-1.0}$ & $3.9\pm1.9$         & -- & $110\pm40$ & $3.1\pm0.9$ & $1.2\pm0.4$ \\
      & strong dip & $36\pm15$ & $8\pm8$  & --          & $1.5\pm1.2$ & $3.1^{+0.9}_{-1.0}$ & $4.9\pm0.9$         & $3.7\pm2.0$         & -- & $60\pm40$ & $1.6\pm1.0$ & $0.6\pm0.4$ \\
     \hline
    \multirow{4}{*}{\fbox{8525}} 
      & non-dip     & $55\pm16$       & $33\pm7$   & --                  & $1.5\pm1.1$ & $2.1\pm1.0$ & $2.4\pm0.9$ & --                  & -- & $90\pm40$ & $2.8\pm1.0$ & $1.0\pm0.4$ \\
      & weak dip   & $68\pm17$       & $26\pm8$   & $6\pm4$             & $2.1\pm1.2$ & $4.6\pm0.9$ & $2.7\pm1.0$ & --                  & -- & $110\pm40$ & $3.1\pm1.1$ & $1.2\pm0.5$ \\
      & dip        & $53\pm16$       & $24\pm8$   & $8.4^{+2.9}_{-3.0}$ & $2.6\pm1.1$ & $3.9\pm0.9$ & $4.4\pm0.9$ & $4.6^{+1.8}_{-1.9}$ & -- & $100\pm40$ & $2.9\pm1.1$ & $1.1\pm0.4$ \\
      & strong dip & $40^{+16}_{-17}$   & $21\pm8$ & --                  & --          & $2.1\pm1.0$ & $4.0\pm1.0$ & $6.2^{+1.8}_{-1.9}$ & -- & $70\pm40$ & $2.1\pm1.1$ & $0.8\pm0.4$ \\
    \hline
    \multirow{4}{*}{\fbox{9847}} 
      & non-dip     & $56\pm14$      & $14\pm8$          & --      & $1.5\pm1.1$ & $1.2\pm0.9$ & $1.5\pm0.9$ & --          & --      & $70\pm40$ & $2.1\pm0.9$ & $0.8\pm0.4$ \\
      & weak dip   & --            & $33\pm9$          & --      & $3.2\pm1.3$ & $4.0\pm1.1$ & $3.4\pm1.1$ & --          & --      & $44\pm19$ & $1.2\pm0.6$ & $0.49\pm0.21$ \\
      & dip        & $51^{+18}_{-19}$ & $19^{+9}_{-10}$ & --      & $3.6\pm1.3$ & $3.8\pm1.1$ & $5.2\pm1.0$ & $5.8\pm2.1$ & --      & $90\pm50$ & $2.5\pm1.2$ & $1.0\pm0.5$ \\
      & strong dip & $42\pm19$     & $11\pm10$         & $4\pm4$ & $3.6\pm1.3$ & $5.5\pm1.0$ & $5.1\pm1.1$ & $7.1\pm2.2$ & $8\pm6$ & $90\pm50$ & $2.5\pm1.3$ & $1.0\pm0.5$ \\
    \hline
\end{tabular}}\\%
{{}$^{\dagger}$\,\citet{herrero95} provide a scaling factor for the
  observed overabundance of helium, which we apply to the solar
  abundances of \citet{wilms00}. See text for further details.}
\end{table*}

Summing the column densities for all ions of an element then yields a
lower limit for the total column density of the element. For comparison
with the continuum fitting values, the abundance of the element can
then be utilized to convert this column density into a corresponding
$N_\mathrm{H}$ value. The values listed in Tables~\ref{tab:N_si}
and~\ref{tab:N_s} list two $N_\mathrm{H}$ values. One value is based
on the solar abundances as summarized by \citet{wilms00}. In our
second conversion we consider that in their analysis of the optical
spectrum of \hde \citet{herrero95} found that helium is overabundant
by a factor of 2.53 with respect to the solar value. Assuming that
metals scale with the same factor as helium with respect to hydrogen,
we scale the solar abundances of \citet{wilms00} by this factor and
use these corrected abundances to derive the corresponding hydrogen
column.

Figure~\ref{fig:dipstage_column} shows how the single ion column
densities vary with depth of the dip. As we enter the dip, for the
lowest ionization states (Fig.~\ref{fig:dipstage_column}, bottom
panels) the columns increase toward the deepest dip stages. This
trend reverses for the highest ionization states
(Fig.~\ref{fig:dipstage_column}, top panels). This behavior is
also indicative of an absorber with a layered ionization structure, i.e., a
medium whose outer regions are more highly ionized than the central
core region.

For the non-dip data, column densities of H-like and He-like silicon
have already been presented in paper~I. Despite the fact that our dip
selection criterion for the non-dip data has not changed, there is a
slight difference in the equivalent width determined for He-like
\ion{Si}{xiii}, where paper~I finds an equivalent width of
$-1.73$\,eV, while we find $-1.92$\,eV, resulting in a $\sim$10\%
difference in the column. Both values are still in agreement within
their uncertainties. The difference in equivalent width for H-like
\ion{Si}{xiv}, on the other hand, is a surprising factor 1.3. A
possible explanation is that the non-dip data allowed a global fit,
including interstellar absorption, more than a hundred lines, and a
pileup correction. The three dip stages, however, do not have enough
signal to constrain such a model, and consequently for consistency we
also model the non-dip data with local fits. Differences at the 15\%
level would therefore be expected owing to these different analysis
approaches. Taking into account these additional 15\%, the H-like
\ion{Si}{xiv} columns are also consistent within their uncertainties.

\begin{figure*}
  \sidecaption
  \includegraphics[width=12cm]{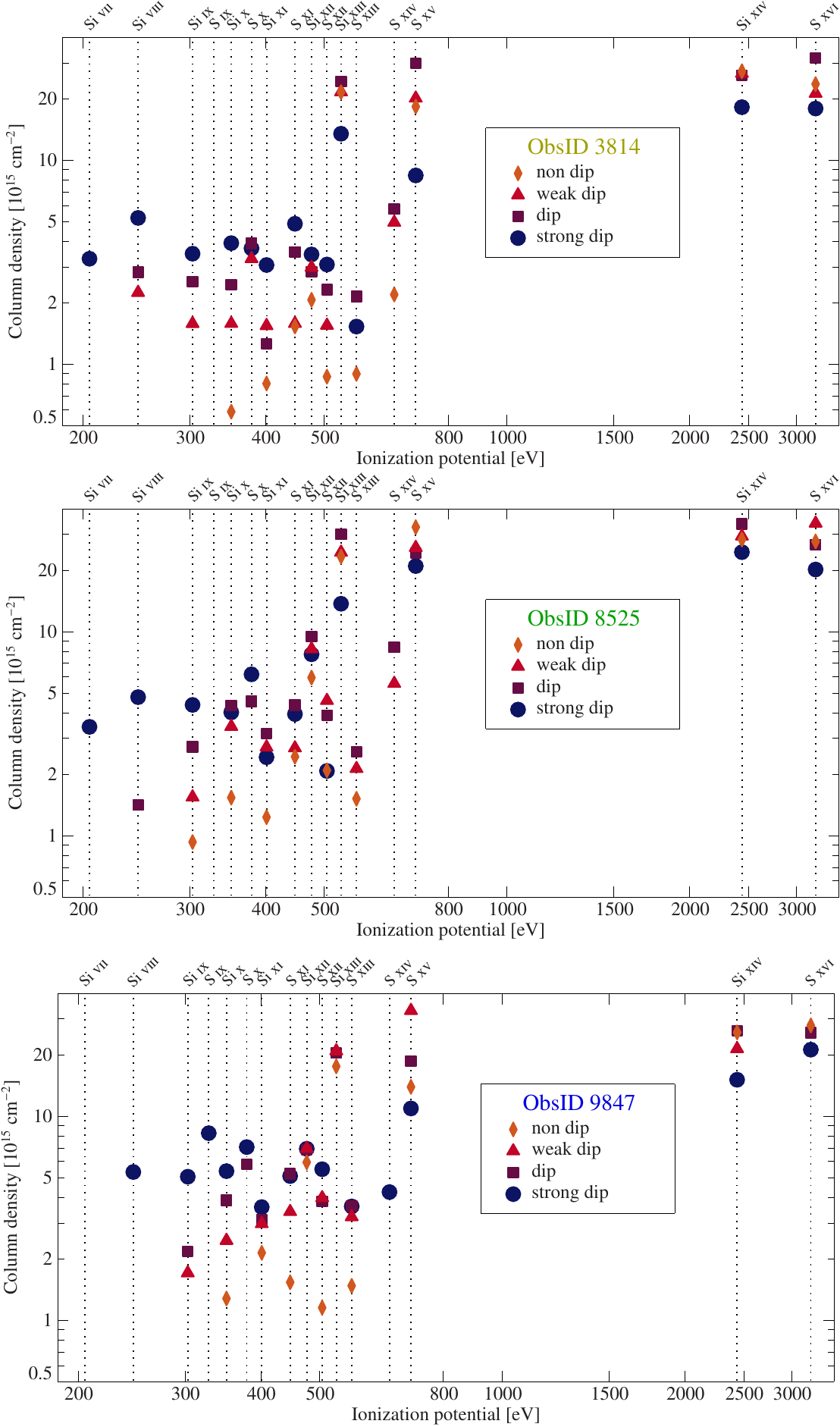}
  \caption{Column densities vs. ionization potential for the
    respective silicon and sulfur ions in all dip stages.}
  \label{fig:ionpot_column}
\end{figure*}

Figure~\ref{fig:ionpot_column} shows the column densities as a
function of the ionization potential of the respective ions. The
closed shell ions He-like \ion{Si}{xiii} and \ion{S}{xv} exhibit
enhanced column densities as expected. However, whereas the appearance
of lower ionization stages in the strong dip points to a lower
temperature in the core of a clump, it is also possible that we see a
constant temperature clump with a higher density in its core. As the
absorber is ionized, the column density measured from continuum
absorption is only a lower limit; we cannot see the fully ionized
material. The figure also clearly illustrates the large column of the
hydrogenic lines. It is therefore very likely that a significant
amount of the S and Si in the system are fully ionized and thus cannot
be detected (see also Sect.~\ref{sec:clumpyabs}).

\subsubsection{Doppler shifts}\label{sec:doppler}

To analyze the morphology of the absorbing material, we next take a
look at the bulk motion, measuring the Doppler shifts of the lines in
the Si and S regions for each dipping stage. Since the uncertainty of
the theoretical rest-wavelengths is on the order of the expected line
shifts \citep{hell13a}, we measured the centroids of the line blends
for each ion in the laboratory using an electron beam ion trap and a
microcalorimeter \citep{hell16}. For the shifts of the He- and H-like
$1\mathrm{s}\rightarrow2\mathrm{p}$ transitions, we used the tables of
\citet{drake88a} and \citet{garcia65a}, respectively, as reference.
These are the same reference publications that we also used for the
calibration of the laboratory data.

\begin{figure*}
  \begin{minipage}[t]{\columnwidth}
    \resizebox{\hsize}{!}{\includegraphics{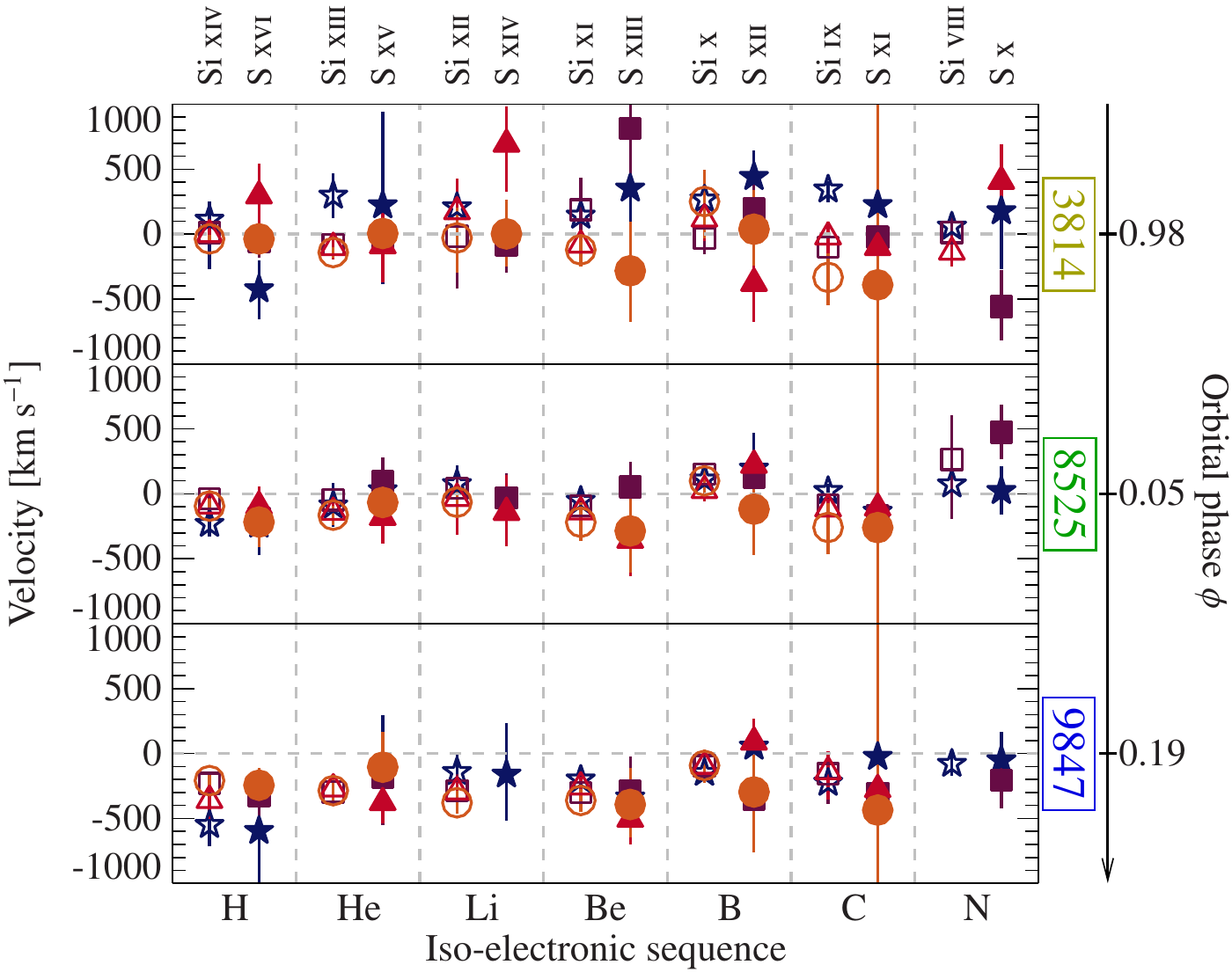}}
    \caption{Doppler velocities for the various ionization states of
      \ion{Si}{viii} -- \ion{Si}{xiv} (empty) and \ion{S}{x} --
      \ion{S}{xvi} (filled) in all dipping stages (non-dip: orange
      circles; weak dip: red triangles; dip: violet squares; and strong dip:
      blue stars) over all observations in the different orbital phases.}
    \label{fig:sishifts}
  \end{minipage}
  \hfill
  \begin{minipage}[t]{\columnwidth}
    \resizebox{\hsize}{!}{\includegraphics{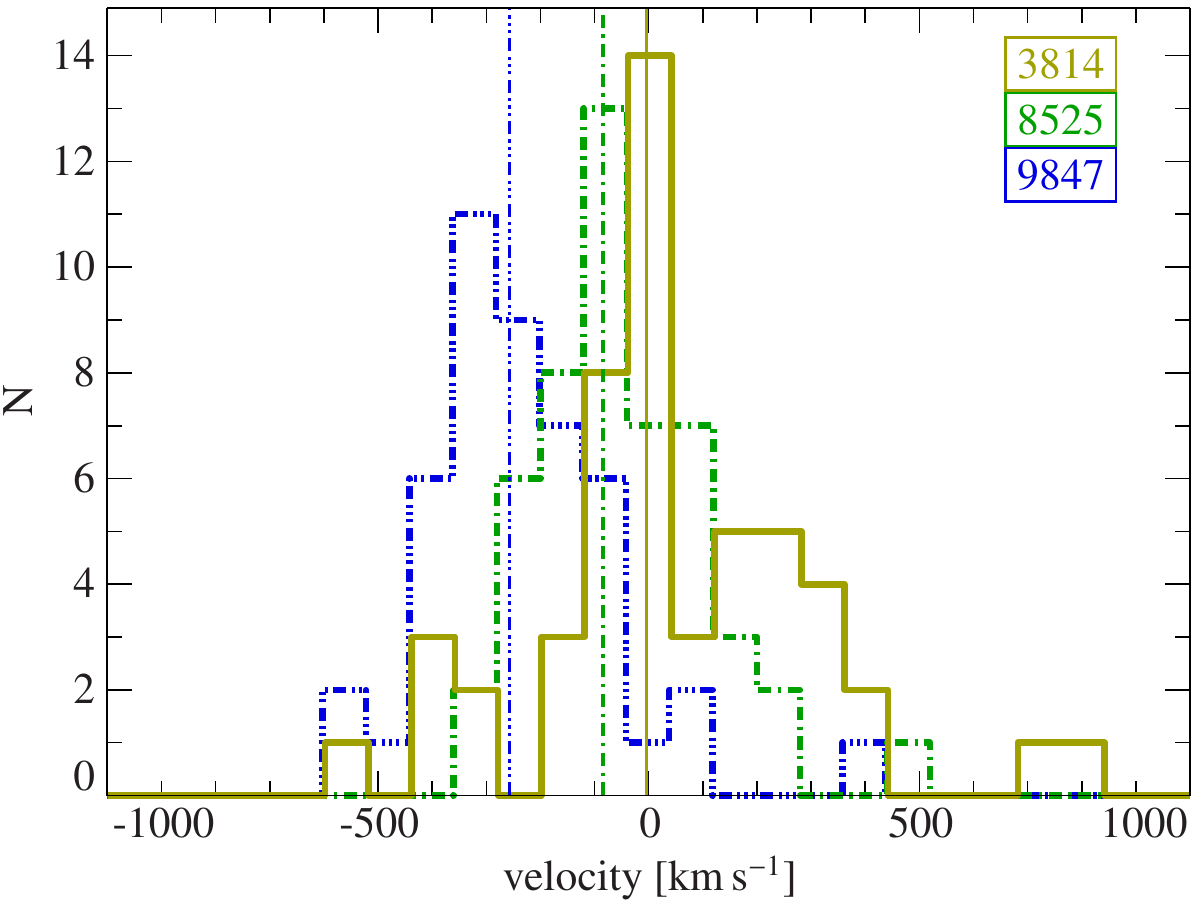}}
    \caption{Distribution of velocity shifts of all Si and S lines
      detected in the three ObsIDs with their medians. The peak velocity
      clearly shifts between the observations.} 
    \label{fig:vhisto}
  \end{minipage}
\end{figure*}

Figure~\ref{fig:sishifts} shows the resulting velocities for each line
in the different dipping stages of each observation. Overall, for the
same spectral lines the Doppler shifts are consistent throughout the
different dipping stages. This is more obvious for the uncontaminated
lines in the Si spectra, where statistics are much better than for the
S lines and line centers can be determined more accurately. The O-like
\ion{S}{ix} line detected in ObsID 9847 is the only outlier. As the
width of this line is narrow (only one energy bin), and since we only
claim a tentative detection of this line (see Sect.~\ref{sec:lines}),
the derived Doppler velocity is probably inaccurate and we do not
include the line in Fig.~\ref{fig:sishifts}.

Figure~\ref{fig:vhisto} shows the distribution of the Doppler shifts
in the three observations. For ObsID 3814, the velocities scatter
around $0\,\mathrm{km}\,\mathrm{s}^{-1}$. For ObsID 8525, the
velocities increase toward negative values and scatter around
$-60\,\mathrm{km}\,\mathrm{s}^{-1}$, and for ObsID 9847, the scatter
moves down to around $-230\,\mathrm{km}\,\mathrm{s}^{-1}$.

This Doppler shift cannot be due to the free fall velocity of donor
material onto the black hole. As we look at the system near
$\phiorb = 0$, the free fall velocity would cause a redshift, not a
blueshift of lines as seen in the spectra. The Keplerian velocity of
the black hole cannot be the cause for the energy shift either, as we
would expect the maximum amplitude of the shift to be on the order of
$100\,\mathrm{km}\,\mathrm{s}^{-1}$ and not much larger than that;
\citep[the projected semi-amplitude of the black hole is
$\mathrm{K}_{\mathrm{BH}} \sim 91\,\mathrm{km}\,\mathrm{s}^{-1}$
(e.g.,][]{brocksopp99, gies03, orosz11}.

\subsubsection{A clumpy absorber}\label{sec:clumpyabs}
\begin{figure}
  \centering
  \includegraphics[viewport=92 530 334 769,clip,width=0.9\columnwidth]{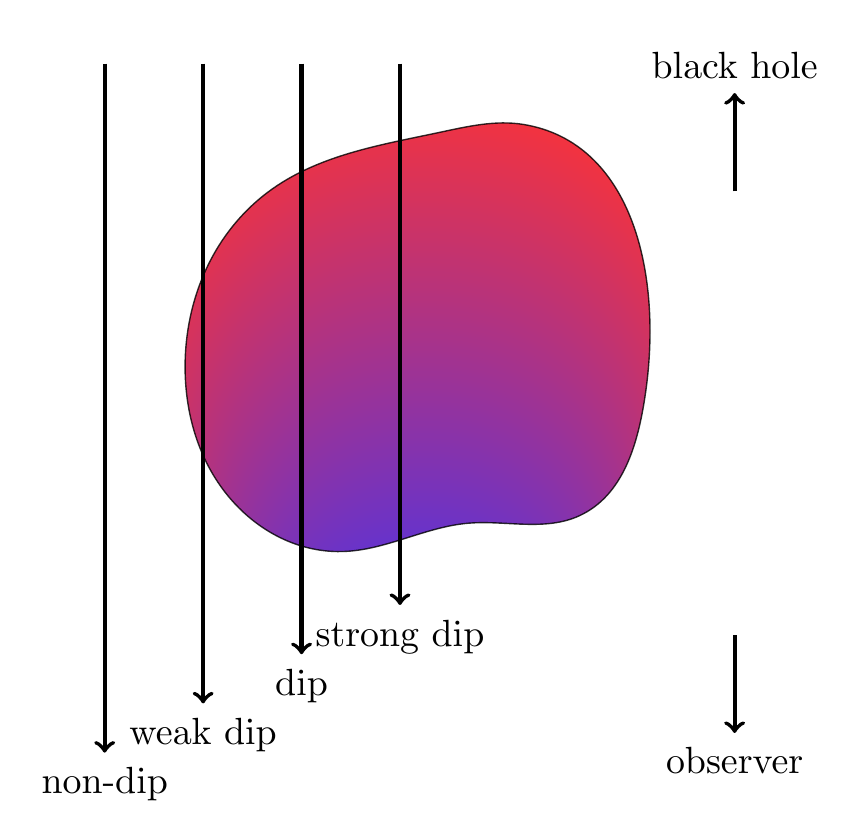}
\caption{Sketch of a clump passing our line of sight and causing the
  different dip stages. Red color indicates more highly ionized
  regions; blue color regions indicate lower ionization balance.}
\label{fig:blob}
\end{figure}

The consistency of the Doppler shifts of the lines during dipping
provides evidence for inhomogeneities in the absorber that are
forming a unified structure that has some kind of density
stratification. We call these structures clumps, even though
we do not want to imply that the structures are static phenomena as
opposed to transient structures. Either owing to self shielding, or
because the material is in some kind of approximate pressure
equilibrium similar to the clouds in active galaxies
\citep{krolik:81a}, the part of the clump facing away from the black
holeis less ionized. As the clump moves through the line of
sight (Fig.~\ref{fig:blob}), we first look through its outer, more
ionized regions. We only see the lower ionization stages during the
deeper dip where the middle of the clump is located in our line of
sight.

\begin{figure}
\resizebox{\hsize}{!}{\includegraphics{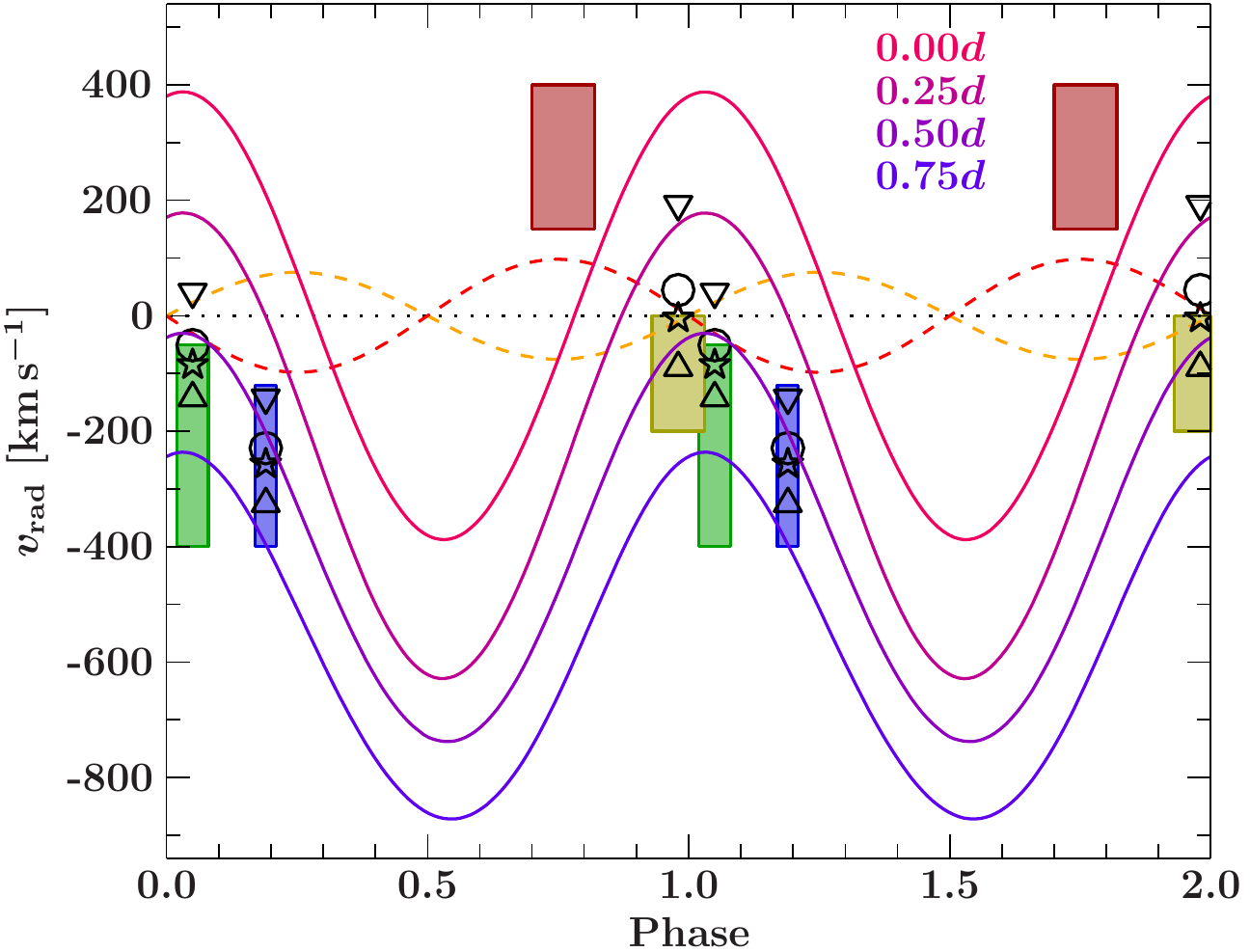}}
\caption{Radial velocities of the black hole (red dashed line) and the
  companion (orange dashed line), and wind velocities projected onto
  the line of sight as a function of distance from the black hole.
  Distances are given in units of the distance between the black hole
  and its donor, $d$. Colored boxes indicate the range of measured
  non-dip Doppler shifts from the \chandra observations (red: ObsID
  3815; yellow: ObsID 3814; green: ObsID 8525; and blue: ObsID 9847,
  modified and corrected from paper~II, Fig.~11). Symbols show the
  median (star), mean (circle), first (point-up triangle), and third
  quartile (point-down triangle) of the distribution of Doppler shifts
  for the Si and S lines.}
\label{fig:ivi}
\end{figure}

Regarding the orbital phase dependency, from the three ObsIDs
analyzed  we can already see a trend toward higher Doppler velocities
further away from \phiorb $\sim 0$. This is consistent with paper~II.
Figure~\ref{fig:ivi} shows the median, mean, first, and third quartile
of the velocity distribution of the Si and S lines in comparison to
the non-dip results shown in Fig.~11 of paper~II. The phase dependence
of the Doppler shifts measured within the dips as well as the
amplitude are roughly consistent with paper~II. The material producing
the dips thus seems to be at the same distance from the black hole as
what can be seen during non-dip.

The projected wind velocities in this figure are calculated using a
CAK-model with a focused wind based on \citet[][see paper~II for more
details\footnote{There was a sign error in the code that produced the
  figure in paper~II. This led to a flip of sign for the velocity
  curves and a slight phase shift. These changes do not have an
  influence on the interpretation of the figure in
  paper~II. Figure~\ref{fig:ivi} shows the corrected velocity
  curves.}]{giesbolton86b}. The model consists of a radially symmetric
wind that is focused onto the black hole in a cone of $20^\circ$.
\citet{giesbolton86b} stated that their model is valid only out to
$0.9d$ away from the stellar center, where $d$ is the distance between
the star and the black hole, as the density at this distance is
already too small to produce a significant contribution to the
absorption line profiles. It is possible to argue that this is only
the case for a smooth wind and clumps have to be treated differently.
Even for $\phiorb=0$ the regions we are interested in (distances of
$0$--$0.75d$ from the black hole, in the direction of the line of
sight at different phases) are mostly outside of the valid region of
the model of \citet{giesbolton86b}, but as our observations are near
$\phiorb=0$, where the distance to the star is not much more than
$0.9d$, we still take the model for a rough estimate. However, both
measured non-dip and dip Doppler shifts fit neither the projected wind
velocity (difference in phase) nor the black hole (difference in
amplitude). Thus, we cannot make an exact statement on the location of
the clumps in the system from our data, although a comparison of the
amplitude implies a distance of less than $0.25\,d$ from the black
hole.

\begin{figure}
  \resizebox{\hsize}{!}{\includegraphics{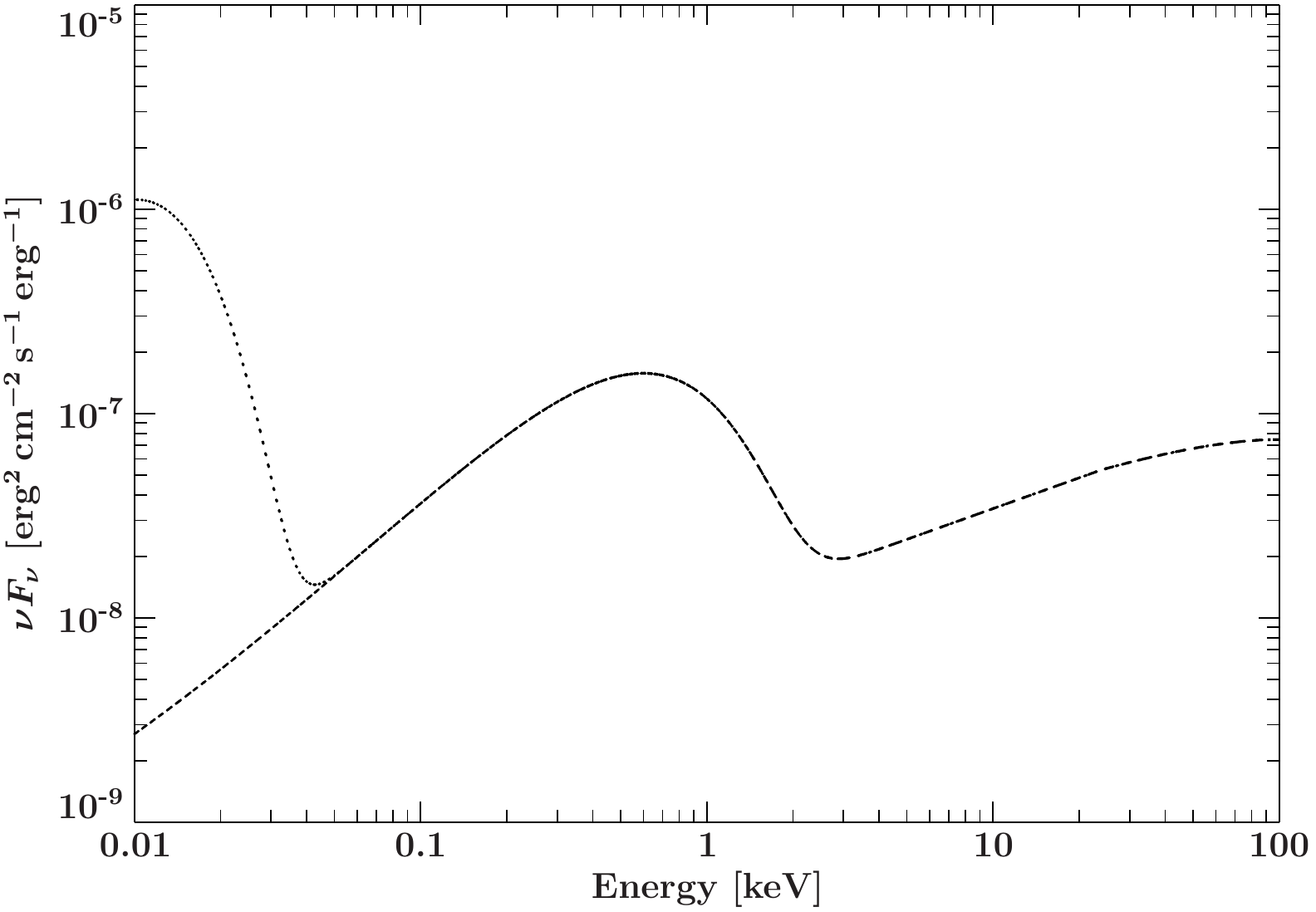}}
  \caption{Incident SED used for the XSTAR simulations. The dotted
    and dashed lines show the SED with and without the optical
    companion star, respectively. We note that the SED is always
    renormalized to the assumed model luminosity.}
  \label{fig:input_sed}
\end{figure}

In order to investigate the origin of the observed ionization, we use
XSTAR \citep{Kallman2001}, which models the photoionization structure
of spherically symmetric gas clouds that are irradiated by a central
source of arbitrary spectral energy distribution. We assume gas clouds
with the default elemental composition used in XSTAR, which is based
on \citet{Grevesse1996}. The clouds have a covering fraction of 100\%,
a temperature of $10^6\,\mathrm{K}$, a constant total hydrogen
particle density of $n=10^{10}\,\mathrm{cm}^{-3}$, and are irradiated
by an incident 1--1000\,Ry luminosity of
$L=1.4\times10^{37}\,\mathrm{erg}\,\mathrm{s}^{-1}$. In this
configuration, the definition of the ionization parameter is
$\xi = L/\left(n R^2\right)$, where $R$ the inner radius of the shell.
Our incident spectral energy density (SED) is adapted from
\citet{Pepe2015} and assumes that the source of X-rays is point like.
Any mechanical input from the jet in the system is ignored. In order
to gauge the influence of the strong UV photon field of the donor
star, we perform simulations with and without a stellar contribution
to the total SED (see Fig.~\ref{fig:input_sed}).

\begin{figure*}
  \includegraphics[height=25\baselineskip]{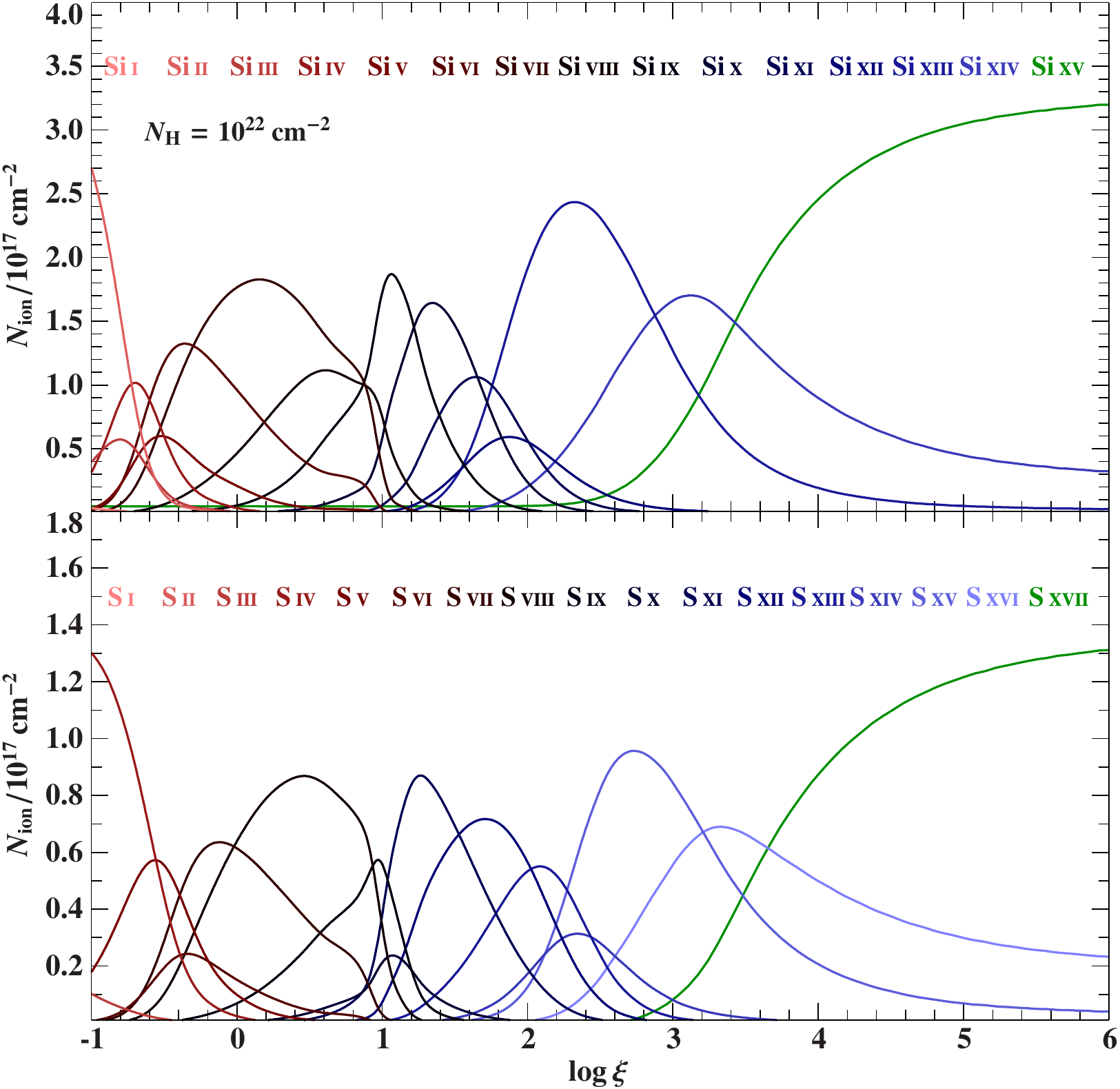}\hfill
  \includegraphics[height=25\baselineskip]{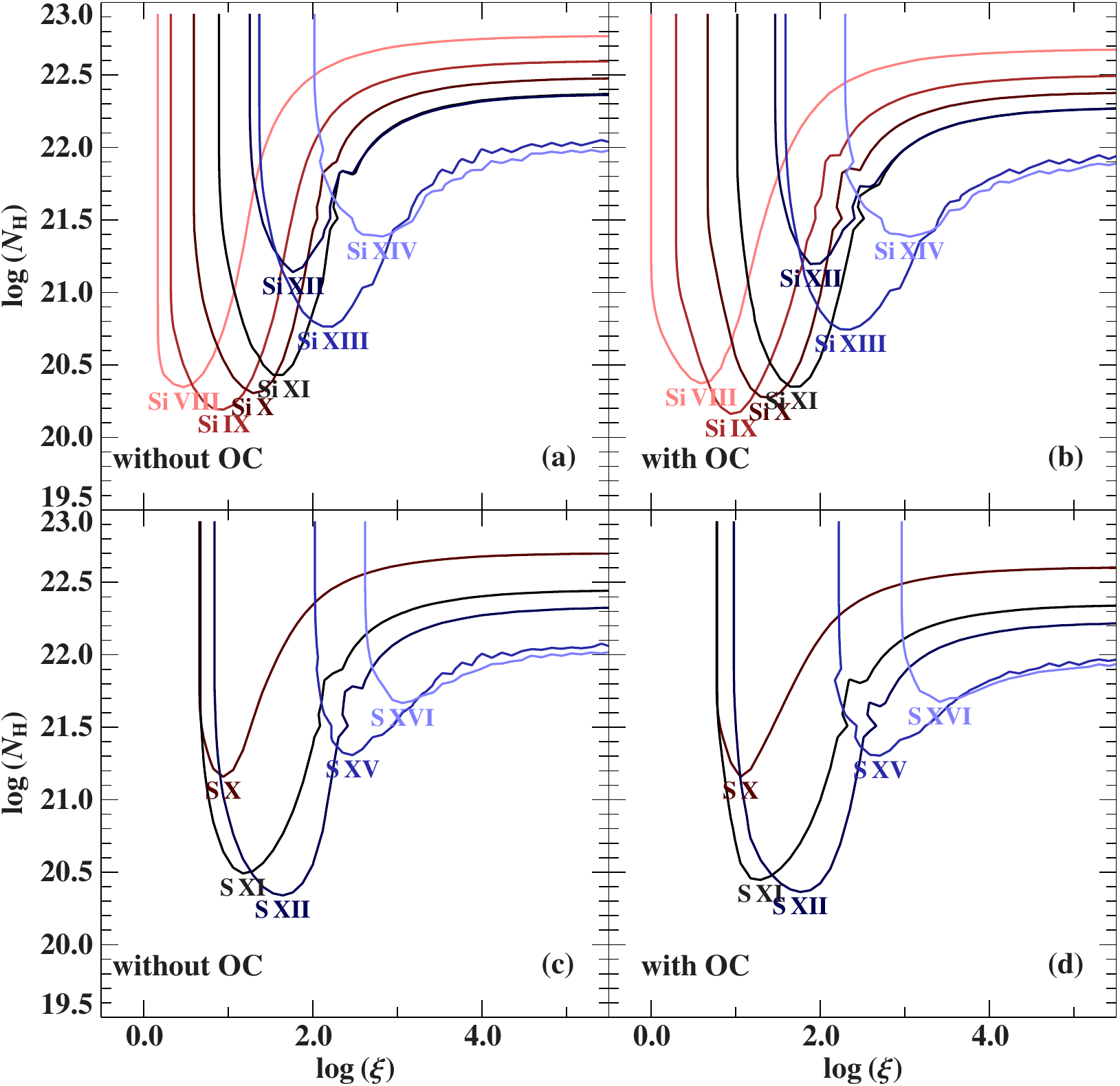}

  \caption{\emph{Left:} Predicted columns for the ions of Si and S for
    a spherical cloud with $N_\mathrm{H}=10^{22}\,\mathrm{cm}^{-2}$
    that is irradiated by a X-ray source with the spectrum defined by
    Fig.~\ref{fig:input_sed}. \emph{Right:} Lines of constant ion
    column density color-coded for each different ion of Si (panels a
    and b) and S (panels c and d), for XSTAR simulated values of
    equivalent hydrogen column density, $N_\mathrm{H}$, and ionization
    parameter, $\log\xi$. The assumed ion columns are the values
    measured during the strong dip of observation 8525
    (Table~\ref{tab:eqws}). If a simple XSTAR model described the
    observation, we would expect these lines to cross in one point
    (within their error bars). The left column of the figure shows
    calculations without the effect of the optical companion (OC); the
    right column includes the UV radiation from the
    companion.}\label{fig:ioncols}
\end{figure*}

We calculate model grids by varying $\xi$ and the total hydrogen
column, $N_\mathrm{H}= n \ell$, where $\ell$ is the geometric size of
the cloud. Since $L$ and $n$ are fixed, this approach effectively
models photoionized clouds of varying sizes and distances from the
source of ionizing photons. The left panel of Fig.~\ref{fig:ioncols}
shows the columns for the ions of Si and S as a function of $\log\xi$
for $N_\mathrm{H}=10^{22}\,\mathrm{cm}^{-2}$. There is no single value
of $\log\xi$ that produces the measured columns for all ions under
consideration. This result holds for all values of
$N_\mathrm{H}$. Based on a grid of XSTAR simulations in which we vary
$\log\xi$ and $N_\mathrm{H}$, in the right-hand panel of
Fig.~\ref{fig:ioncols}, we show the contours in the
$\log\xi$-$N_\mathrm{H}$-plane in which the theoretical column of each
ion of Si and S ion equals that measured during the strong dip of
observation 8525. In the ideal case of a simple medium of constant
density and temperature, all of these contours would intersect in a
single point; the modulo measurement errors are, however, small
compared to the large dynamic range shown in Fig.~\ref{fig:ioncols}.
We note that the inclusion of the optical companion has only a
marginal effect on the observed ion column densities of Si and S
(Fig.~\ref{fig:ioncols}, right, panels c and~d). This is important
because the binary geometry of the system with two sources of
radiation cannot be accounted for by the XSTAR simulation. Neglecting
the optical companion therefore only introduces a small systematic
effect on ionization structure modeling of these ions.

In practice, assuming that the dip is produced by a single absorbing
cloud, we would expect some kind of a density variation in the cloud,
as discussed above. We can use our simple XSTAR simulations to
identify regions of particular interest. From Fig.~\ref{fig:ioncols},
columns of moderately ionized ions can be roughly reproduced with
hydrogen column densities around $3\times10^{20}\,\mathrm{cm}^{-2}$
and ionization parameters of $\log\xi{\sim}1$--2. The high ionization
states of silicon and sulfur, on the other hand, require column
densities and an ionization parameter that are roughly an order of
magnitude higher. This result points at a more complex origin for the
ionization structure than photoionization alone. One possible
explanation would be ionization due to the strong shocks that are
present in hydrodynamic simulations for massive X-ray binary systems
\citep[][and references
therein]{blondin90,blondin91,blondin94,manousakis:11a,manousakis2015,sundqvist2018}
and in winds from early-type stars \citep[][and references
therein]{owocki1988,sundqvist2013}.

\subsubsection{Line symmetry}
\begin{figure*}
\centering
\includegraphics[width=\textwidth]{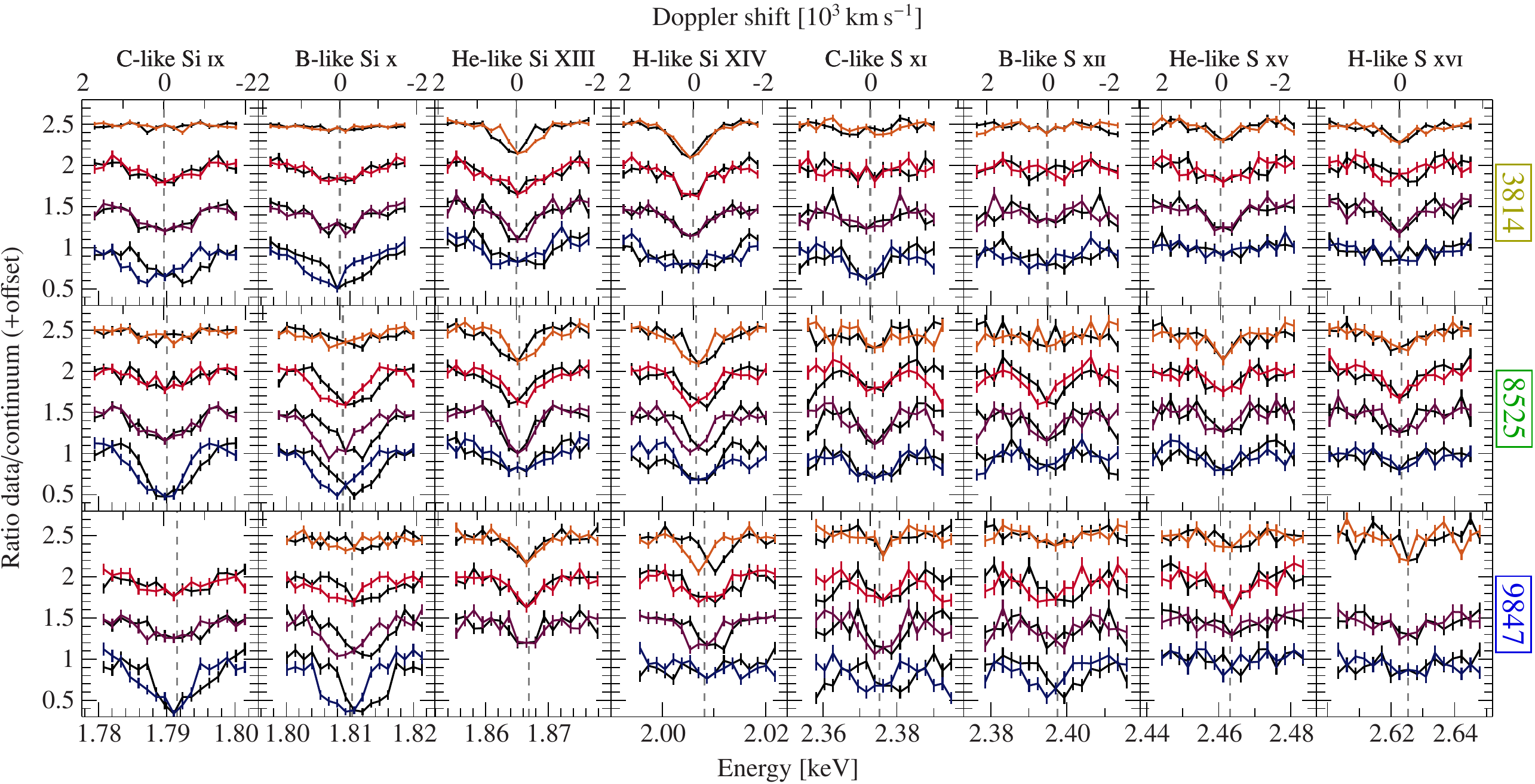}

\caption{Original line profiles (colored) for the non-dip (top,
  orange), weak dip (second from top, red), dip (third from top,
  purple), and strong dip (bottom, blue) lines for selected ions, as
  well as the same data flipped (black, see also text for further
  information) at the energy corresponding to adding a constant
  Doppler shift for each observation (vertical dashed line) to the
  line's laboratory energy. For an ideal, symmetric line, the colored
  and black curves should match. Most of the lines observed have
  complex line profiles.}
\label{fig:symmetry}
\end{figure*}

Although a detailed analysis of the complex photoionization line
profiles observed in Cyg X-1 is beyond the scope of this paper, we can
at least take a quantitative look at the line shapes in the region of
$\pm 8$ bins around the respective rest energy, adjusted for the most
probable Doppler shift for each ObsID (3814:
$0\,\,\mathrm{km}\,\mathrm{s}^{-1}$, 8525:
$-80\,\mathrm{km}\,\mathrm{s}^{-1}$, 9847:
$-320\,\mathrm{km}\,\mathrm{s}^{-1}$; see also Fig.~\ref{fig:vhisto}).
Figure~\ref{fig:symmetry} shows the original data (colored) and the
same data flipped (black) around the center of the bin containing the
adjusted rest energy (gray dashed line). For a symmetric line shape,
the original and flipped data should match, as is the case, for example, for
8525 He-like \ion{Si}{xiii} in the dip data, or 3814 H-like
\ion{Si}{xiv} non-dip, weak dip and dip data. However, often we see the
lines shifted by one or even two bins (e.g., 3814 C-like \ion{Si}{ix}
strong dip or 9847 B-like \ion{S}{xii} weak dip, dip and strong
dip). There are also signs of asymmetry (e.g., 9847 C-like
\ion{Si}{ix} strong dip or 3814 C-like \ion{S}{xi} dip), as well as
lines that show both asymmetry and a shift (e.g., 3814 C-like
\ion{Si}{ix} strong dip or 8525 B-like \ion{Si}{x} non-dip and dip).

Clear P Cygni-like profiles for \cyg were reported only at
$\phiorb=0.5$ during ObsID 11044
\citep{miskovicova11}. \citet{schulz02} also found indications of P
Cygni profiles at $\phiorb=0.74$. In the UV, P Cygni profiles were
found by \citet{vrtilek:08a}, also at $\phiorb=0.5$, where the stellar
wind is focused toward the observer by the black hole. At around
$\phiorb=0$, the focused wind is moving away from the observer toward
the black hole, and the redshift and ionization are
highest. Asymmetries in the lines are thus expected at that phase.
This was observed, for example, by \citet{feng2003} and discussed in
paper~II. Our observations fit this picture.

\subsection{Emission in the deepest dips}

While the absorption lines discussed so far show a clear dependency on
the depth of the dip, this is not the case for the He-like
\ion{Si}{xiii}\,z emission line, whose total line flux is constant
throughout the dip stages. This result could either indicate that the
emitting material originates in a geometrically much larger region
than the absorber, such that the emission line is not affected by the
absorber, or, alternatively, that the emission line originates in an
area where the line of sight does not pass through the absorber.

A possible origin for the \ion{Si}{xiii} emission line could be the
photoionization region in the stellar wind that surrounds the X-ray
emission region. As discussed in paper~I, the high ionization region
around the X-ray source, similar to the Str\"omgren sphere around
stars, is comparable in size to the separation of the black hole and
its donor star, i.e., it is much larger than the absorbing cloud.
Given the low optical depth of this region we expect a region whose
X-ray emission is dominated by emission lines, similar to that seen,
for example, in Vela~X-1
\citep{schulz_vela,watanabe2006,grinberg2017a} or in Cen~X-3
\citep{wojdowski03a}. The flux from the photoionization region,
however, is much fainter than the X-rays from the accretion flow. The
region has therefore been mainly studied for edge-on systems, where
our line of sight onto the bright X-ray source is blocked by the donor
star during eclipses. While the Cyg~X-1/HDE 226868 system does not
show eclipses, we can utilize the strong dipping to at least partially
block the line of sight to the black hole. The best opportunity for
this is presented by the long, deep dip at the beginning ObsID~8525,
which is the most pronounced of all the dips in our observations. The
deep part of this dip (according to our dip stage identification,
i.e., the blue data points in Fig.~\ref{fig:lightcurves} belonging to
that dip) lasted for 5.3\,ks (1.5\,hours), which is long enough to
accumulate a decent gratings spectrum of one single dip. Because of
the strong absorption, pileup is no concern in the resulting spectrum
and the HEG and MEG spectra agree with each other.

\begin{figure}
\resizebox{\hsize}{!}{\includegraphics{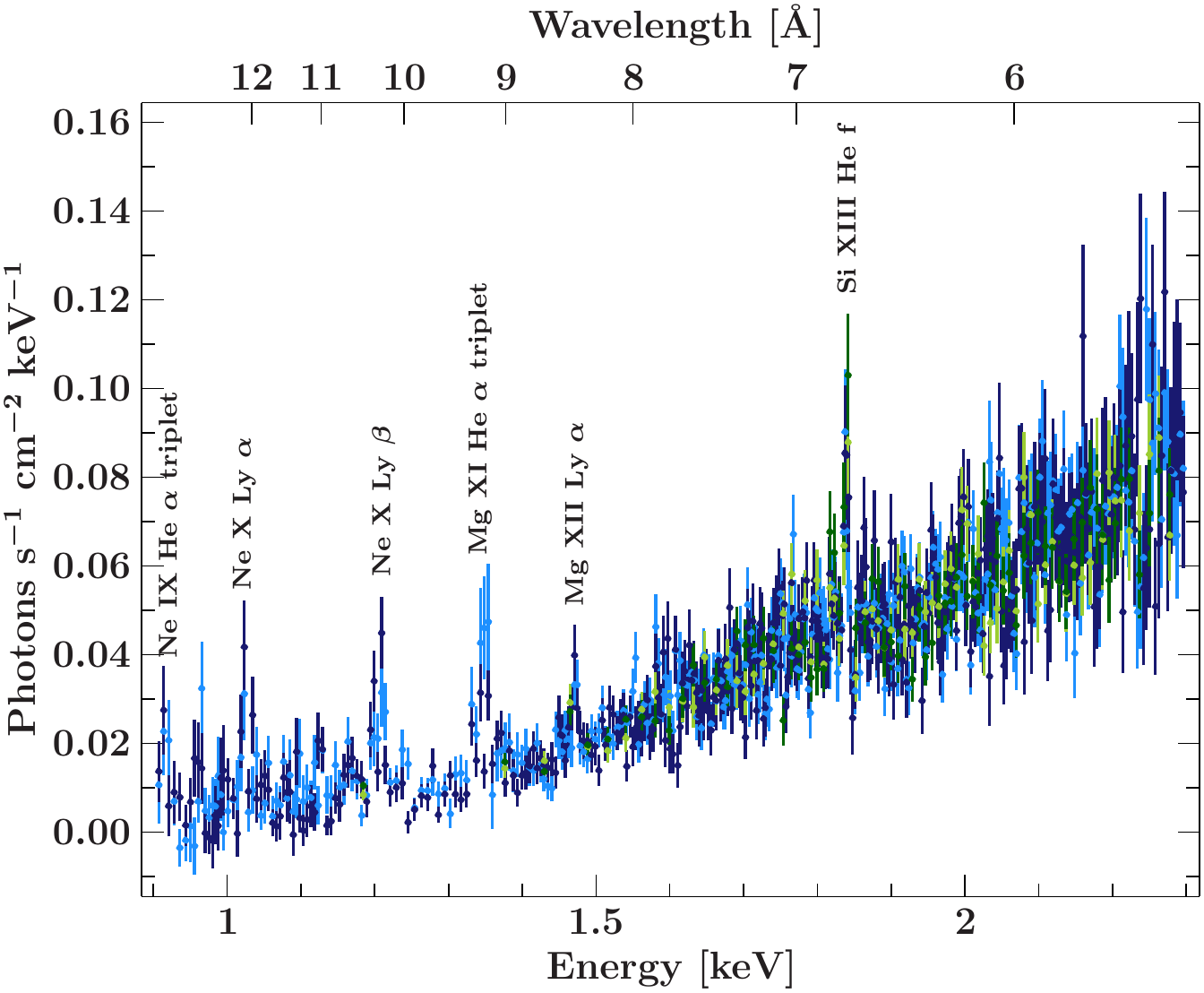}}
\caption{ Deep dip spectrum from ObsID~8525 from a subset of the
  strong dip data in the 0.9--2.3\,keV (13.8--5.4\,{\AA}) range.
The data are rebinned to $S/N \geq 5$, revealing emission lines from the
  photoionized zone around the black hole.}\label{fig:emission}
\end{figure}

As shown in Fig.~\ref{fig:emission}, during the deepest parts of the
dip in ObsID~8525, the photoionized zone around the black hole indeed
dominates the spectrum, resulting in the presence of emission lines.
The most prominent lines detected are \ion{Si}{xiii}\,He\,z
\citep[1.8394\,keV $\equiv$ 6.7405;][]{porquet10a},
Mg\,Ly\,$\alpha$ (1.4723\,keV $\equiv$ 8.4211\,{\AA}), the
Mg\,He\,$\alpha$ triplet ($\sim$1.3522\,keV $\equiv$ 9.1691\,{\AA}),
as well as Ne\,Ly\,$\beta$ (1.2109\,keV $\equiv$ 10.2390\,{\AA}),
Ne\,Ly\,$\alpha$ (1.0280\,keV $\equiv$ 12.0601\,{\AA}), and the
\ion{Ne}{ix}\,He\,$\alpha$ triplet ($\sim$0.9220\,keV $\equiv$
13.4473\,{\AA}). Unfortunately, however, the signal to noise in the
lines is too low to allow a further characterization of their
parameters.

\section{Summary}\label{sec:summary}

We have analyzed three \chandra-HETGS observations of \cyg around
superior conjunction of the system, where the structure of the stellar
wind causes absorption features, so-called dips, in the X-ray
light curve. With the help of color-color diagrams, we divide the data
into four different dip stages containing a comparable total number of
counts. For each dip stage, we fit the silicon and sulfur regions of
the spectrum by applying a local power law continuum and Gaussian lines
to model the absorption lines for different ionization stages of
silicon and sulfur. We find lower charge ions appearing in the deeper
dip stages, equivalent widths / absorption columns rising for deeper
dip stages, phase dependent Doppler shifts for the lines, asymmetric
line shapes, and emission lines emerging in the deepest dip.

The lower charge states of both elements only appear in the deeper dip
stages, pointing toward colder material shielded from the irradiation
by the black hole. The total columns in the dip stages derived from
the equivalent widths of the lines represent lower limits to the real
values, as we cannot see the fully ionized material adding to the
column, which is likely to carry most of the mass. The derived
$N_\mathrm{H}$ values of
$N_\mathrm{H} \sim 2\dots 4 \cdot 10^{21}\,\mathrm{cm}^{-2}$ are in
agreement with previous results from paper~I, although there are
slight differences due to differences in the continuum.

The Doppler shifts of the lines with respect to measured laboratory
values \citep{hell16} show a scatter between the different dip stages, but the same trend within a single observation, pointing toward a
single structure containing different ionization stages, which moves
as a whole with a certain speed. The Doppler velocities show a clear
modulation with orbital phase and are in agreement with the results
found in paper~II. However, these velocities match neither the expected wind
velocities calculated from a toy model after \citet{giesbolton86a} nor
free fall nor the Kepler velocity around the black hole. Thus, from
the line shifts it is not possible to tell where the material causing
the absorption dips in the light curve is located.

For the three observations around $\phiorb=0$, the lines show
asymmetries. This is expected from previous observations near
superior conjunction (see paper~I and references therein). Clear P
Cygni-like profiles are reported only near $\phiorb=0.5$.

Within the deepest dip, there is an emission line spectrum emerging
that is outshone otherwise. Together with the \ion{Si}{xiii} emission
line that does not change its total line flux during the dip stages,
this emission line spectrum indicates a much larger area responsible
for the emission than what is absorbed by the wind structures, or an
area distant from the absorber such that the line of sight toward it
is not affected. A possibility is the almost fully ionized
photoionization region in the stellar wind surrounding the X-ray
emission region.

\section{Conclusions}
Our analysis gives us a glimpse at the properties of the structures
causing the dips. From the light curve, we can constrain the duration
of a dip, and thus of a passage of the clump through our line of
sight, from 0.5\,ks for the shorter dips up to 5\,ks for the longer
dips. This assumption is only valid if a dip is caused by a single,
bigger clump with a cold or dense core instead of many small clumps in
the line of sight.

Assuming a distance of $0.25\,d$ between the clump and the black hole
($d$ being the distance between black hole and donor; see
Sect.\,\ref{sec:doppler}), a simple Keplerian approach (i.e., the
clump orbiting the black hole) gives a
velocity of ${\sim}460\,\mathrm{km}\,\mathrm{s}^{-1}$ perpendicular to
the line of sight. From the velocity of the clump and the duration of
the dip, we can estimate the diameter of the clumps to
$(2.3\ldots23)\times 10^{8}$\,m, equivalent to
$(0.33\ldots3.3)\,R_\odot$ or $(0.02\ldots0.2)\,R_*$. As an advancing
clump grows in size \citep{sundqvist:12}, we can calculate its
ejection size \citep{grinberg2015} to about a third of the size that
we measure at the distance of $0.25\,d$.

With the current data we can only speculate whether this picture of
single, big clumps is the correct one. The morphology of the light
curve in ObsID 8525 supports this idea, whereas ObsID~3814 with its
rapid variability appears to be more in favor of a large number of
small clumps in our line of sight. For a sufficiently high number of
smaller clumps, those closer to the black hole are partly shielding
the outer ones from the ionizing radiation, which is in agreement with
our observation of lower ionization stages during strong dipping. A
possible explanation would then be a mixture of some larger clumps,
causing the distinct dips in the light curve, and many smaller clumps.

This idea is consistent with the picture of structured, radiation
pressure driven O-star winds \citep{castor75}. As shown, for example,
by \citet{OwockiRybicki1984}, the stationary solution for a
line-driven wind is unstable. These instabilities grow quickly and
result in strong shocks. Nonstationary hydrodynamic simulations show
that dense cool shells of gas already form in deep wind regions close
to the photosphere of the O star
\citep{feldmeier1997,DessartOwocki2003,oskinova2012,sundqvist2018}.
Large density, velocity, and temperature variations due to the
de-shadowing instability further compress the gas in these shells and
fragment these shells into clumps. Theory predicts thus that the wind
is actually a two-phase medium consisting of tenuous, hot gas
(covering most of the volume) and embedded cool, dense clumps
(containing most of the mass). Observational evidence for this
clumping has been found for isolated O-type supergiants
\citep[e.g.,][]{Eversberg1998,Markova2005,Bouret2005,Fullerton2006,Oskinova2006},
but also for high mass X-ray binaries \citep[e.g.,][]{torrejon2015}.

Speculatively, the picture of the dip behavior would then be that
observations at phase~0, which sample the region of the stellar wind
closer to the donor star, should show a wide spectrum of clump sizes.
Observations at phase 0.75 sample regions that are farther away from
the wind, where the clump size distribution has changed, probably
favoring larger clumps, resulting in less short -term $N_\mathrm{H}$
variation. Clearly, however, using only three observations from
different orbits and because of the need to perform joint spectroscopy of
multiple dips, we do not have sufficient statistics to make a firm
statement about the distribution and size of the clumps. Our estimates
of $(0.02\dots 0.2)\mathrm{R}_{*}$ for the diameter of a clump are in
agreement with the high end of the recent 2D simulation results of
\citet{sundqvist2018}, however, who have found a typical length scale for
clumps of $~0.01\mathrm{R}_{*}$ at a distance of two stellar radii
from the star. We cannot distinguish between quasi-spherical clumps
and, for instance, pancake-shaped clumps as proposed by \citet{oskinova2012},
which would have different optical depths for different lines of
sight. \citet{sundqvist2018} have even found various different clump shapes
coexisting in their 2D simulations. Athena with its
larger effective area will provide more insight as it will be able to
look at single dips.

\begin{acknowledgements}
  We thank the referee for their very constructive report which
  improved the paper.
  
  The research leading to these results was funded by the
  Bundesministerium f\"ur Wirtschaft und Technologie under grant
  numbers \mbox{DLR 50 OR 0701} and \mbox{DLR 50 OR 1113}, by LLNL
  under Contract DE-AC52-07NA27344, and is supported by NASA grants to
  LLNL. VG is supported through the Margarete von Wrangell fellowship
  by the ESF and the Ministry of Science, Research and the Arts
  Baden-Württemberg.

  This research has made use of ISIS functions provided by ECAP/Remeis
  observatory and MIT
  (\url{http://www.sternwarte.uni-erlangen.de/isis/}). We thank John
  E. Davis for providing the \texttt{slxfig} module used for creating
  the presented plots.

  We thank Ivica~Mi\v{s}kovi\v{c}ov\'a for her initial reduction of
  the data sets used here.
\end{acknowledgements}

\begin{table*}\renewcommand{\arraystretch}{1.2}
  \caption{Detector constants and continuum parameters for all
    observations. We note that the power law fits are applied to a very
    local continuum (1.605--2.045\,keV $\equiv$
    7.725--6.063\,{\AA} for the Si region and 2.295--2.7\,keV
    $\equiv$ 5.402--4.6\,{\AA} for the S region). The detector
    constant accounts for the different flux normalizations of the HEG
    and the MEG.} 
  \label{tab:continuum}
  \begin{tabular}{lllll}
    \hline
    \hline
    \multirow{2}{*}{\fbox{3814}} &  & detector constant & power law norm & photon index \\
                              &  &          & \tiny{[photons\,keV$^{-1}$\,cm$^{-2}$\,s$^{-1}$]} & \\
    \hline
    Si region 
      & non-dip     & $0.779\pm0.005$           & $0.709^{+0.017}_{-0.016}$  & $1.12\pm0.04$ \\
      & weak dip   & $0.830\pm0.010$           & $0.493^{+0.024}_{-0.023}$  & $0.84\pm0.08$ \\
      & dip        & $0.873\pm0.010$           & $0.306^{+0.015}_{-0.014}$  & $0.43\pm0.08$ \\
      & strong dip & $0.958^{+0.014}_{-0.013}$ & $0.057\pm0.004$            & $-0.83\pm0.10$ \\
    \cline{2-5}
    S region  
      & non-dip     & $0.948\pm0.010$           & $1.14^{+0.13}_{-0.12}$    & $1.60\pm0.12$ \\
      & weak dip   & $0.966\pm0.018$           & $0.87^{+0.22}_{-0.18}$    & $1.42\pm0.25$ \\
      & dip        & $0.985^{+0.018}_{-0.017}$ & $0.74^{+0.18}_{-0.15}$    & $1.37\pm0.23$ \\
      & strong dip & $1.010\pm0.018$           & $0.088^{+0.023}_{-0.018}$ & $-0.31\pm0.25$ \\
    \hline
    \hline
    \multirow{2}{*}{\fbox{8525}} &  & detector constant & power law norm & photon index \\
                              &  &          & \tiny{[photons\,keV$^{-1}$\,cm$^{-2}$\,s$^{-1}$]} & \tiny{$[\times10^{-4}]$}\\
    \hline
    Si region 
      & non-dip     & $0.791\pm0.009$           & $0.63\pm0.04$                & $0.95\pm0.09$ \\
      & weak dip   & $0.832^{+0.011}_{-0.010}$ & $0.457^{+0.025}_{-0.023}$    & $0.70\pm0.09$ \\
      & dip        & $0.900\pm0.012$           & $0.247^{+0.014}_{-0.013}$    & $0.20\pm0.09$ \\
      & strong dip & $0.989\pm0.016$           & $0.0368^{+0.0025}_{-0.0023}$ & $-1.27\pm0.11$ \\
    \cline{2-5}
    S region  
      & non-dip     & $1.000^{+0.022}_{-0.021}$ & $1.04^{+0.25}_{-0.20}$    & $1.45\pm0.23$ \\
      & weak dip   & $0.960\pm0.020$           & $0.56^{+0.17}_{-0.13}$    & $0.87\pm0.28$ \\
      & dip        & $0.987\pm0.019$           & $0.44^{+0.12}_{-0.10}$    & $0.84\pm0.26$ \\
      & strong dip & $1.022^{+0.020}_{-0.019}$ & $0.082^{+0.025}_{-0.017}$ & $-0.33^{+0.29}_{-0.26}$ \\
    \hline
    \hline
    \multirow{2}{*}{\fbox{9847}} &  & detector constant & power law norm & photon index \\
                              &  &          & \tiny{[photons\,keV$^{-1}$\,cm$^{-2}$\,s$^{-1}$]} & \tiny{$[\times10^{-4}]$}\\
    \hline
    Si region 
      & non-dip     & $0.728\pm0.008$ & $0.84\pm0.04$             & $1.08\pm0.08$ \\
      &  weak dip  & $0.773\pm0.012$ & $0.62\pm0.04$             & $0.84\pm0.11$ \\
      &  dip       & $0.846\pm0.013$ & $0.374^{+0.026}_{-0.024}$ & $0.43^{+0.12}_{-0.11}$ \\
      & strong dip & $0.970\pm0.018$ & $0.062\pm0.005$           & $-1.12\pm0.12$ \\
    \cline{2-5}
    S region  
      & non-dip     & $0.925\pm0.019$           & $1.29^{+0.26}_{-0.27}$ & $1.50^{+0.20}_{-0.25}$ \\
      &  weak dip  & $0.958^{+0.030}_{-0.029}$ & $1.4^{+0.5}_{-0.4}$    & $1.74^{+0.29}_{-0.30}$ \\
      &  dip       & $0.978^{+0.029}_{-0.028}$ & $0.64^{+0.20}_{-0.16}$ & $1.01^{+0.29}_{-0.30}$ \\
      & strong dip & $1.003^{+0.030}_{-0.029}$ & $0.16^{+0.06}_{-0.04}$ & $0.1\pm0.4$ \\
    \hline
\end{tabular}
\end{table*}

\begin{table*}\renewcommand{\arraystretch}{1.2}
  \caption{Observed silicon and sulfur line centers for all observations.}
  \label{tab:energies}
  \begin{tabular}{lllllll}
    \hline
    \hline
    \multirow{2}{*}{\fbox{3814}} & \multicolumn{2}{c}{\multirow{2}{*}{Ion}} & \multicolumn{4}{c}{$E_\mathrm{obs}$\,[keV]} \\ 
    \cline{4-7}
    & & & non-dip & weak dip & dip & strong dip \\ 
    \hline
    & H-like  & \ion{Si}{xiv}  & $2.00574(23)$ & $2.0055(5)$  & $2.0054(6)$  & $2.0048(26)$  \\
    & He-like & \ion{Si}{xiii} & $1.86588(22)$ & $1.8656(6)$  & $1.8655(4)$  & $1.8632(11)$  \\
    & Li-like & \ion{Si}{xii}  & $1.8452(17)$  & $1.8440(28)$ & $1.8451(26)$ & $1.8437(14)$  \\
    & Be-like & \ion{Si}{xi}   & $1.8283(8)$   & $1.8280(7)$  & $1.8264(19)$ & $1.8267(7)$   \\
    & B-like  & \ion{Si}{x}    & $1.8069(17)$  & $1.8077(11)$ & $1.8086(8)$  & $1.80678(43)$ \\
    & C-like  & \ion{Si}{ix}   & $1.7916(13)$  & $1.7897(8)$  & $1.7902(9)$  & $1.7875(6)$   \\
    & N-like  & \ion{Si}{viii} & --            & $1.7728(8)$  & $1.7719(7)$  & $1.7717(5)$   \\
    & O-like  & \ion{Si}{vii}  & --            & --           & --           & $1.7571(15)$  \\
    \cline{2-7}
    & H-like  & \ion{S}{xvi}  & $2.6220(9)$  & $2.6191(22)$ & $2.6222(11)$ & $2.6254(20)$ \\
    & He-like & \ion{S}{xv}   & $2.4606(9)$  & $2.4614(24)$ & $2.4608(12)$ & $2.459(6)$   \\
    & Li-like & \ion{S}{xiv}  & $2.4378(22)$ & $2.4321(30)$ & $2.4385(19)$ & --           \\
    & Be-like & \ion{S}{xiii} & $2.420(4)$   & --           & $2.4110(88)$ & $2.4147(29)$ \\
    & B-like  & \ion{S}{xii}  & $2.3947(25)$ & $2.3980(24)$ & $2.3934(21)$ & $2.3914(19)$ \\
    & C-like  & \ion{S}{xi}   & $2.3759(16)$ & $2.374(7)$   & $2.3730(18)$ & $2.3710(11)$ \\
    & N-like  & \ion{S}{x}    & --           & $2.3472(23)$ & $2.3548(22)$ & $2.3490(35)$ \\
    & O-like  & \ion{S}{ix}   & --           & --           & --           & --           \\
    \hline
    \hline
    \multirow{2}{*}{\fbox{8525}} & \multicolumn{2}{c}{\multirow{2}{*}{Ion}} & \multicolumn{4}{c}{$E_\mathrm{obs}$\,[keV]} \\ 
    \cline{4-7}
    & & & non-dip & weak dip & dip & strong dip \\ 
    \hline
    & H-like  & \ion{Si}{xiv}  & $2.0061(5)$  & $2.0061(5)$   & $2.0057(4)$   & $2.0070(7)$   \\
    & He-like & \ion{Si}{xiii} & $1.8660(5)$  & $1.8659(5)$   & $1.8653(4)$   & $1.8656(11)$  \\
    & Li-like & \ion{Si}{xii}  & $1.8454(8)$  & $1.8452(18)$  & $1.8448(6)$   & $1.8446(9)$   \\
    & Be-like & \ion{Si}{xi}   & $1.8289(10)$ & $1.8284(5)$   & $1.8281(4)$   & $1.8278(6)$   \\
    & B-like  & \ion{Si}{x}    & $1.8078(9)$  & $1.80823(47)$ & $1.80748(31)$ & $1.80772(41)$ \\
    & C-like  & \ion{Si}{ix}   & $1.7911(13)$ & $1.7903(9)$   & $1.7901(7)$   & $1.78944(38)$ \\
    & N-like  & \ion{Si}{viii} & --           & --            & $1.7705(27)$  & $1.7716(6)$   \\
    & O-like  & \ion{Si}{vii}  & --           & --            & --            & $1.7568(21)$  \\
    \cline{2-7}
    & H-like  & \ion{S}{xvi}  & $2.6236(17)$ & $2.6225(13)$ & $2.6234(18)$ & $2.6238(22)$ \\
    & He-like & \ion{S}{xv}   & $2.4612(11)$ & $2.4621(17)$ & $2.4598(15)$ & $2.4604(19)$ \\
    & Li-like & \ion{S}{xiv}  & --           & $2.4389(24)$ & $2.4380(17)$ & --           \\
    & Be-like & \ion{S}{xiii} & $2.4198(27)$ & $2.4204(23)$ & $2.4171(16)$ & --           \\
    & B-like  & \ion{S}{xii}  & $2.3959(29)$ & $2.3932(9)$  & $2.3940(11)$ & $2.3934(23)$ \\
    & C-like  & \ion{S}{xi}   & $2.3749(17)$ & $2.3737(18)$ & $2.3742(10)$ & $2.3739(20)$ \\
    & N-like  & \ion{S}{x}    & --           & --           & $2.3467(17)$ & $2.3502(15)$ \\
    & O-like  & \ion{S}{ix}   & --           & --           & --           & --           \\
    \hline
    \hline
    \multirow{2}{*}{\fbox{9847}} & \multicolumn{2}{c}{\multirow{2}{*}{Ion}} & \multicolumn{4}{c}{$E_\mathrm{obs}$\,[keV]} \\ 
    \cline{4-7}
    & & & non-dip & weak dip & dip & strong dip \\ 
    \hline
    & H-like  & \ion{Si}{xiv}  & $2.0069(4)$ & $2.0079(10)$ & $2.0070(7)$  & $2.0092(12)$  \\
    & He-like & \ion{Si}{xiii} & $1.8668(6)$ & $1.8667(6)$  & $1.8668(7)$  & --            \\
    & Li-like & \ion{Si}{xii}  & $1.8474(6)$ & $1.8468(10)$ & $1.8468(9)$  & $1.8459(10)$  \\
    & Be-like & \ion{Si}{xi}   & $1.8297(6)$ & $1.8291(5)$  & $1.8293(6)$  & $1.8287(5)$   \\
    & B-like  & \ion{Si}{x}    & $1.8089(9)$ & $1.8090(8)$  & $1.8090(5)$  & $1.80928(37)$ \\
    & C-like  & \ion{Si}{ix}   & --          & $1.7904(13)$ & $1.7905(14)$ & $1.79092(37)$ \\
    & N-like  & \ion{Si}{viii} & --          & --           & --           & $1.7725(5)$   \\
    & O-like  & \ion{Si}{vii}  & --          & --           & --           & --            \\
    \cline{2-7}
    & H-like  & \ion{S}{xvi}  & $2.6238(12)$ & --           & $2.6246(19)$ & $2.6269(39)$ \\
    & He-like & \ion{S}{xv}   & $2.4615(23)$ & $2.4637(14)$ & $2.4621(30)$ & $2.4618(36)$ \\
    & Li-like & \ion{S}{xiv}  & --           & --           & --           & $2.4391(33)$ \\
    & Be-like & \ion{S}{xiii} & $2.4207(23)$ & $2.4216(16)$ & $2.4198(25)$ & $2.4202(17)$ \\
    & B-like  & \ion{S}{xii}  & $2.397(4)$   & $2.3942(16)$ & $2.3978(15)$ & $2.3945(11)$ \\
    & C-like  & \ion{S}{xi}   & $2.3763(20)$ & $2.3750(18)$ & $2.3752(11)$ & $2.3730(14)$ \\
    & N-like  & \ion{S}{x}    & --           & --           & $2.3520(17)$ & $2.3508(22)$ \\
    & O-like  & \ion{S}{ix}   & --           & --           & --           & $2.3291(26)$ \\
    \hline
  \end{tabular}
\end{table*}

\begin{table*}
\caption{Evolution of the equivalent widths of the silicon and sulfur
  lines during the dipping stages for all observations.}\label{tab:eqws}
\begin{tabular}{lllllll}
  \hline
  \hline
  \multirow{2}{*}{\fbox{3814}} & \multicolumn{2}{c}{\multirow{2}{*}{Ion}} & \multicolumn{4}{c}{Equivalent width [eV]}\\
  \cline{4-7}
  & & & non-dip & weak dip & dip & strong dip \\
  \hline
  & H-like & \ion{Si}{xiv} & $-2.48\pm0.14$ & $-2.42^{+0.29}_{-0.28}$ & $-2.38^{+0.29}_{-0.28}$ & $-1.7\pm0.4$ \\
  & He-like & \ion{Si}{xiii} & $-1.92\pm0.14$ & $-1.93\pm0.28$ & $-2.16\pm0.27$ & $-1.2\pm0.4$ \\
  & Li-like & \ion{Si}{xii} & $-0.46\pm0.16$ & $-0.7^{+0.4}_{-1.0}$ & $-0.6\pm0.4$ & $-0.8^{+0.5}_{-0.6}$ \\
  & Be-like & \ion{Si}{xi} & $-0.49\pm0.13$ & $-0.95\pm0.25$ & $-0.77^{+0.27}_{-0.26}$ & $-1.88\pm0.28$ \\
  & B-like & \ion{Si}{x} & $-0.42\pm0.14$ & $-1.13^{+0.26}_{-0.25}$ & $-1.76\pm0.24$ & $-2.81\pm0.26$ \\
  & C-like & \ion{Si}{ix} & $-0.28\pm0.14$ & $-1.13\pm0.25$ & $-1.80^{+0.25}_{-0.24}$ & $-2.48^{+0.27}_{-0.26}$ \\
  & N-like & \ion{Si}{viii} & -- & $-0.83\pm0.26$ & $-1.05\pm0.26$ & $-1.94\pm0.29$ \\
  & O-like & \ion{Si}{vii} & -- & -- & -- & $-0.5\pm0.4$ \\
  \cline{2-7}
  & H-like & \ion{S}{xvi} & $-2.2\pm0.4$ & $-1.9\pm0.8$ & $-2.9\pm0.7$ & $-1.6\pm0.7$ \\
  & He-like & \ion{S}{xv} & $-1.6\pm0.4$ & $-1.8\pm0.7$ & $-2.7\pm0.6$ & $-0.8\pm0.7$ \\
  & Li-like & \ion{S}{xiv} & $-0.5\pm0.4$ & $-1.1\pm0.7$ & $-1.3\pm0.7$ & -- \\
  & Be-like & \ion{S}{xiii} & $-0.6\pm0.4$ & -- & $-1.3\pm0.7$ & $-1.0\pm0.7$ \\
  & B-like & \ion{S}{xii} & $-0.6\pm0.4$ & $-1.1\pm0.7$ & $-1.7\pm0.7$ & $-2.3\pm0.7$ \\
  & C-like & \ion{S}{xi} & $-1.1\pm0.4$ & $-1.2\pm0.9$ & $-2.6\pm0.7$ & $-3.6\pm0.7$ \\
  & N-like & \ion{S}{x} & -- & $-1.3\pm0.8$ & $-1.5\pm0.8$ & $-1.4\pm0.8$ \\
  & O-like & \ion{S}{ix} & -- & -- & -- & -- \\
  \hline
  \hline
  \multirow{2}{*}{\fbox{8525}} & \multicolumn{2}{c}{\multirow{2}{*}{Ion}} & \multicolumn{4}{c}{Equivalent width [eV]}\\
  \cline{4-7}
  & & & non-dip & weak dip & dip & strong dip \\
  \hline
  & H-like & \ion{Si}{xiv} & $-2.58^{+0.29}_{-0.28}$ & $-2.67\pm0.30$ & $-3.07\pm0.28$ & $-2.2\pm0.4$ \\
  & He-like & \ion{Si}{xiii} & $-2.08\pm0.28$ & $-2.18\pm0.29$ & $-2.67\pm0.28$ & $-1.2\pm0.4$ \\
  & Li-like & \ion{Si}{xii} & $-1.3\pm0.4$ & $-1.8\pm0.4$ & $-2.1\pm0.4$ & $-1.7\pm0.4$ \\
  & Be-like & \ion{Si}{xi} & $-0.76\pm0.26$ & $-1.67\pm0.25$ & $-1.94\pm0.25$ & $-1.5\pm0.4$ \\
  & B-like & \ion{Si}{x} & $-1.10\pm0.25$ & $-2.46^{+0.24}_{-0.23}$ & $-3.11^{+0.22}_{-0.21}$ & $-2.88\pm0.28$ \\
  & C-like & \ion{Si}{ix} & $-0.66\pm0.25$ & $-1.10\pm0.26$ & $-1.94\pm0.25$ & $-3.11^{+0.27}_{-0.26}$ \\
  & N-like & \ion{Si}{viii} & -- & -- & $-0.52\pm0.30$ & $-1.8\pm0.4$ \\
  & O-like & \ion{Si}{vii} & -- & -- & -- & $-0.5\pm0.4$ \\
  \cline{2-7}
  & H-like & \ion{S}{xvi} & $-2.5\pm0.8$ & $-3.1\pm0.8$ & $-2.4\pm0.8$ & $-1.8\pm0.8$ \\
  & He-like & \ion{S}{xv} & $-2.9\pm0.7$ & $-2.3\pm0.7$ & $-2.2\pm0.7$ & $-1.9\pm0.7$ \\
  & Li-like & \ion{S}{xiv} & -- & $-1.3^{+0.8}_{-0.7}$ & $-1.9\pm0.7$ & -- \\
  & Be-like & \ion{S}{xiii} & $-0.9\pm0.7$ & $-1.3\pm0.8$ & $-1.6\pm0.7$ & -- \\
  & B-like & \ion{S}{xii} & $-1.5\pm0.7$ & $-3.4\pm0.7$ & $-2.9\pm0.7$ & $-1.5\pm0.8$ \\
  & C-like & \ion{S}{xi} & $-1.8\pm0.7$ & $-2.0\pm0.8$ & $-3.2\pm0.7$ & $-2.9\pm0.7$ \\
  & N-like & \ion{S}{x} & -- & -- & $-1.8^{+0.8}_{-0.7}$ & $-2.4^{+0.8}_{-0.7}$ \\
  & O-like & \ion{S}{ix} & -- & -- & -- & -- \\
  \hline
  \hline
  \multirow{2}{*}{\fbox{9847}} & \multicolumn{2}{c}{\multirow{2}{*}{Ion}} & \multicolumn{4}{c}{Equivalent width [eV]}\\
  \cline{4-7}
  & & & non-dip & weak dip & dip & strong dip \\
  \hline
  & H-like & \ion{Si}{xiv} & $-2.34^{+0.28}_{-0.27}$ & $-1.9\pm0.4$ & $-2.4\pm0.4$ & $-1.4\pm0.5$ \\
  & He-like & \ion{Si}{xiii} & $-1.56\pm0.28$ & $-1.9\pm0.4$ & $-1.8\pm0.4$ & -- \\
  & Li-like & \ion{Si}{xii} & $-1.3\pm0.4$ & $-1.6\pm0.5$ & $-1.5^{+0.4}_{-0.5}$ & $-1.5\pm0.5$ \\
  & Be-like & \ion{Si}{xi} & $-1.31\pm0.24$ & $-1.82\pm0.29$ & $-1.91\pm0.29$ & $-2.2\pm0.4$ \\
  & B-like & \ion{Si}{x} & $-0.92\pm0.24$ & $-1.76\pm0.29$ & $-2.78\pm0.26$ & $-3.86^{+0.27}_{-0.26}$ \\
  & C-like & \ion{Si}{ix} & -- & $-1.2\pm0.4$ & $-1.55^{+0.31}_{-0.30}$ & $-3.60\pm0.28$ \\
  & N-like & \ion{Si}{viii} & -- & -- & -- & $-2.0\pm0.4$ \\
  & O-like & \ion{Si}{vii} & -- & -- & -- & -- \\
  \cline{2-7}
  & H-like & \ion{S}{xvi} & $-2.5\pm0.7$ & -- & $-2.3\pm0.9$ & $-1.9\pm0.9$ \\
  & He-like & \ion{S}{xv} & $-1.2\pm0.7$ & $-2.9\pm0.8$ & $-1.7\pm0.9$ & $-1.0\pm0.9$ \\
  & Li-like & \ion{S}{xiv} & -- & -- & -- & $-1.0\pm0.9$ \\
  & Be-like & \ion{S}{xiii} & $-0.9\pm0.7$ & $-2.0\pm0.8$ & $-2.3\pm0.8$ & $-2.3\pm0.8$ \\
  & B-like & \ion{S}{xii} & $-0.8\pm0.7$ & $-2.9\pm0.8$ & $-2.8\pm0.8$ & $-4.0\pm0.8$ \\
  & C-like & \ion{S}{xi} & $-1.1\pm0.7$ & $-2.5^{+0.9}_{-0.8}$ & $-3.9\pm0.8$ & $-3.8\pm0.8$ \\
  & N-like & \ion{S}{x} & -- & -- & $-2.3^{+0.9}_{-0.8}$ & $-2.7\pm0.9$ \\
  & O-like & \ion{S}{ix} & -- & -- & -- & $-1.4\pm1.0$ \\
  \hline
\end{tabular}
\end{table*}

\end{document}